\documentclass[10pt,twocolumn,letterpaper]{article}

\usepackage[pagenumbers]{cvpr} %

\newif\ifdraft
\draftfalse

\usepackage{makecell}
\usepackage{newunicodechar}
\newunicodechar{ }{\,}

\ifdraft
\newcommand{\sining}[1]{{\color{orange}[\textbf{Sining:} #1]}}
\newcommand{\guan}[1]{{\color{purple}[\textbf{Guan:} #1]}}
\newcommand{\itai}[1]{{\color{magenta}[\textbf{Itai:} #1]}}
\newcommand{\rana}[1]{{\color{cyan}[\textbf{Rana:} #1]}}
\newcommand{\namanh}[1]{{\color{orange}[\textbf{Nam Anh:} #1]}}

\newcommand{\TODO}[1]{\textbf{\color{red}[TODO: #1]}}

\newcommand{\rh}[1]{{\color{cyan}#1}}

\else
\newcommand{\sining}[1]{}
\newcommand{\guan}[1]{}
\newcommand{\itai}[1]{}
\newcommand{\rana}[1]{}
\newcommand{\namanh}[1]{}

\newcommand{\TODO}[1]{}

\newcommand{\rh}[1]{{#1}}

\fi

\newcommand{\ourmethod}{LL3M}

\newcommand{\Acal}{\mathcal{A}}

\usepackage{graphicx}
\usepackage{caption}  

\usepackage{algorithm}
\usepackage{algpseudocode}
\newcommand*\samethanks[1][\value{footnote}]{\footnotemark[#1]}

\definecolor{cvprblue}{rgb}{0.21,0.49,0.74}
\usepackage[pagebackref,breaklinks,colorlinks,allcolors=cvprblue]{hyperref}
\def\paperID{*****} %
\def\confName{CVPR}
\def\confYear{2026}

\title{\ourmethod{}: Large Language 3D Modelers}

\author{Sining Lu\thanks{Authors contributed equally.} \quad Guan Chen\samethanks \quad Nam Anh Dinh \quad Itai Lang \quad Ari Holtzman \quad Rana Hanocka\\ \vspace{2mm}
University of Chicago}

\begin{document}
\twocolumn[{%
\renewcommand\twocolumn[1][]{#1}%
\maketitle

\begin{center}
    \twocolumn[{%
    \renewcommand\twocolumn[1][]{#1}%
    \maketitle
    \vspace{-3.7em}
    \begin{center}
        \includegraphics[width=\linewidth,trim=0 100 0 0 clip]{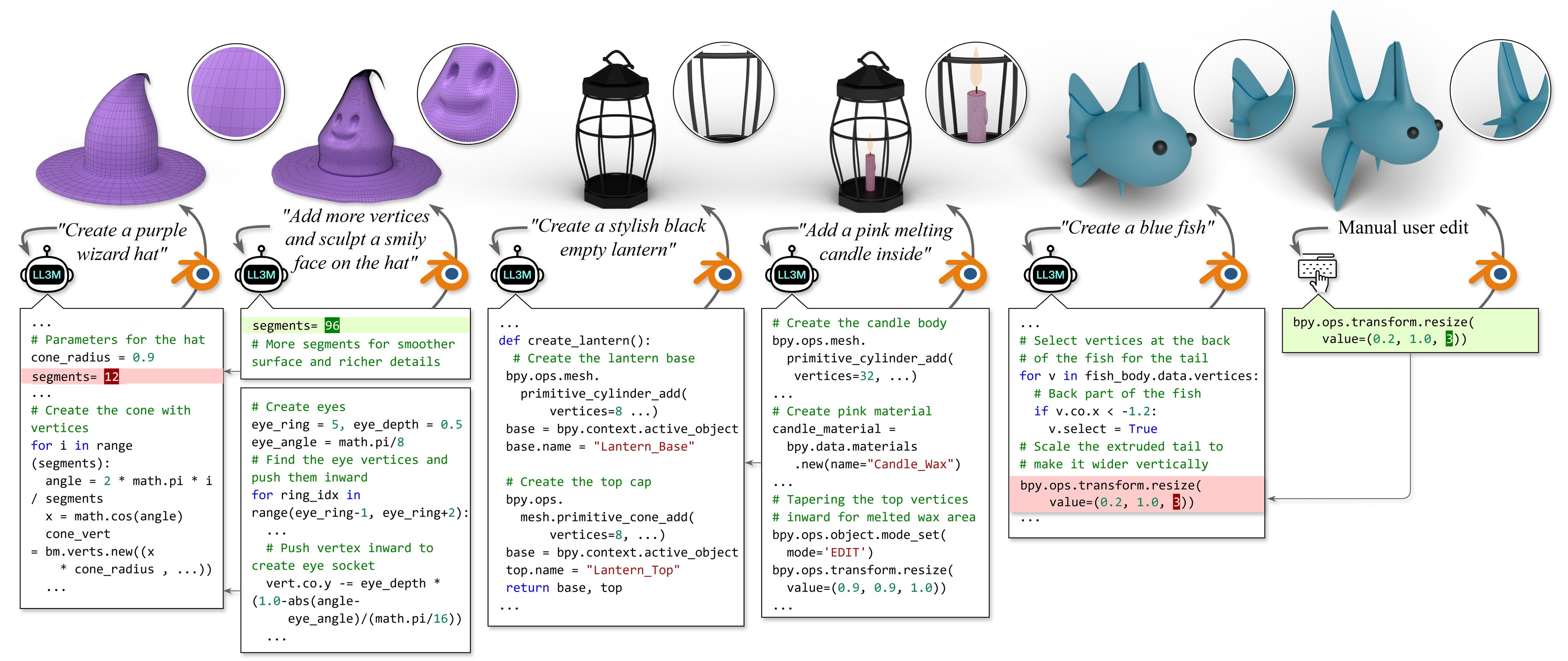}
        \captionof{figure}{\ourmethod{} leverages a team of large language models to write Python code that creates and edits 3D assets in Blender. The agents are capable of creating expressive shapes from scratch and realizing complex, precise geometric manipulations as intuitive code edits. In this example, \ourmethod{} interactively creates and edits 3D assets entirely from user-provided text prompts (see horizontally centered quotes), which can include both creative direction and precise instruction. The agent-generated code is clearly structured and well-documented, making it easy and intuitive for anyone to make changes (see rightmost column, user updates parameter in code to modify object size).}
        \label{fig:teaser}
    \end{center}
    \vspace{1em}
    }]
\end{center}%
}]

\begin{abstract}
We present \ourmethod{}, a multi-agent system that leverages pretrained large language models (LLMs) to generate 3D assets by writing interpretable Python code in Blender. We break away from the typical generative approach that learns from a collection of 3D data. Instead, we reformulate shape generation as a code-writing task, enabling greater modularity, editability, and integration with artist workflows. 
Given a text prompt, \ourmethod{} coordinates a team of specialized LLM agents to plan, retrieve, write, debug, and refine Blender scripts that generate and edit geometry and appearance. The generated code works as a high-level, interpretable, human-readable, well-documented representation of scenes and objects, making full use of sophisticated Blender constructs (e.g. B-meshes, geometry modifiers, shader nodes) for diverse, unconstrained shapes, materials, and scenes. This code presents many avenues for further agent and human editing and experimentation via code tweaks or procedural parameters. This medium naturally enables a co-creative loop in our system: agents can automatically self-critique using code and visuals, while iterative user instructions provide an intuitive way to refine assets. A shared code context across agents enables awareness of previous attempts, and a retrieval-augmented generation knowledge base built from Blender API documentation, BlenderRAG, equips agents with examples, types, and functions empowering advanced modeling operations and code correctness. We demonstrate the effectiveness of \ourmethod{} across diverse shape categories, style and material edits, and user-driven refinements. Our experiments showcase the power of code as a generative and interpretable medium for 3D asset creation. Our project page is at \url{https://threedle.github.io/ll3m}.
\end{abstract}

\section{Introduction}
\label{sec:intro}
The ability to generate editable 3D geometry at scale is becoming increasingly practical in computer graphics and vision. Recent work has explored using various geometric representations for training generative models on large collections of 3D shapes~\cite{siddiqui2024meshgpt,hao2024meshtron,wang2025nautilus}.

\begin{figure}[!t]
    \centering
    \includegraphics[width=\columnwidth]{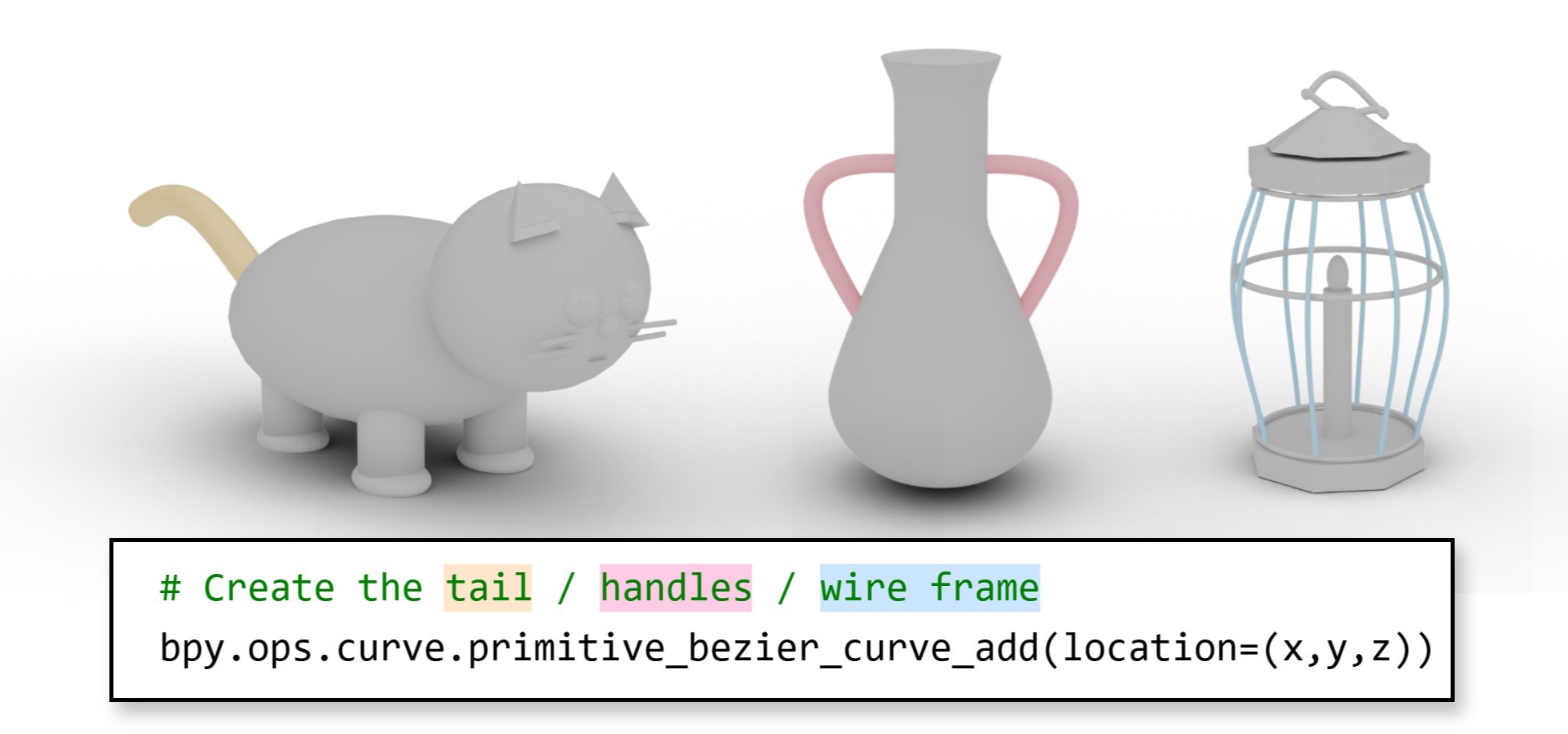}
    \caption{\textbf{Reusable code structure enables generalization.} Despite visual differences, shapes often share high-level code patterns (such as loops, modifiers, and node setups) that recur across categories. This shared structure allows the model to transfer knowledge and generate diverse, editable, and modular code from a wide range of prompts.}
    \label{fig:shared_code_structure}
\end{figure}

We opt to break away from conventional representation-centric generative modeling and ask the question: \emph{Are pretrained large language models (LLMs) capable of generating 3D assets?} A straightforward use of LLMs for mesh generation is to output a list of vertices and faces directly~\cite{taomeshllm2024,wang2023llamamesh}. However, this low-level representation of unordered mesh elements does not effectively capture the high-level structure shared across shapes. 
Inspired by the success of LLMs in generating code~\cite{jiang2024surveylargelanguagemodels, md2024mapcoder, jiang2025ossbenchbenchmarkgeneratorcoding, guo2024deepseekcoderlargelanguagemodel, quan2025codeelobenchmarkingcompetitionlevelcode}, we reformulate shape generation as a \emph{code writing task}. This enables leveraging the abstraction and reusability that code naturally provides across different shapes. For instance, the function calls used to define a Bézier curve for elements like a vase handle, a lamp wire, or chair legs often share similar structure and properties (see \cref{fig:shared_code_structure}). 
\rh{Existing work has made similar observations, and thus leveraged LLMs to generate Blender code: to retrieve existing assets and place them in a scene~\cite{ziniu2024scenecraft}, generate materials for existing 3D assets~\cite{ian2024blenderalchemy}, and generate parameters for predefined procedural scenes~\cite{chunyi20233dgpt}. Unlike these approaches, our method uses LLMs to write Blender code that generates comprehensive, open-vocabulary 3D assets with detailed geometry and appearance attributes.

We introduce \ourmethod{}, which produces complete 3D assets from text prompts without requiring specialized datasets or finetuning.} A straightforward use of an LLM to write Blender scripts leads to code that produces low-quality outputs or fails to execute. To address this, we design a novel multi-agent framework in which a team of LLM agents, specializing in distinct roles, collaborate to guide the 3D generation process. These agents coordinate through structured communication and use multiple modalities and interfaces: planning how to construct the shape, retrieving relevant code snippets, debugging scripts, and rendering outputs for visual evaluation and refinement. 

\begin{figure}[!t]
    \centering
    \includegraphics[width=\columnwidth]{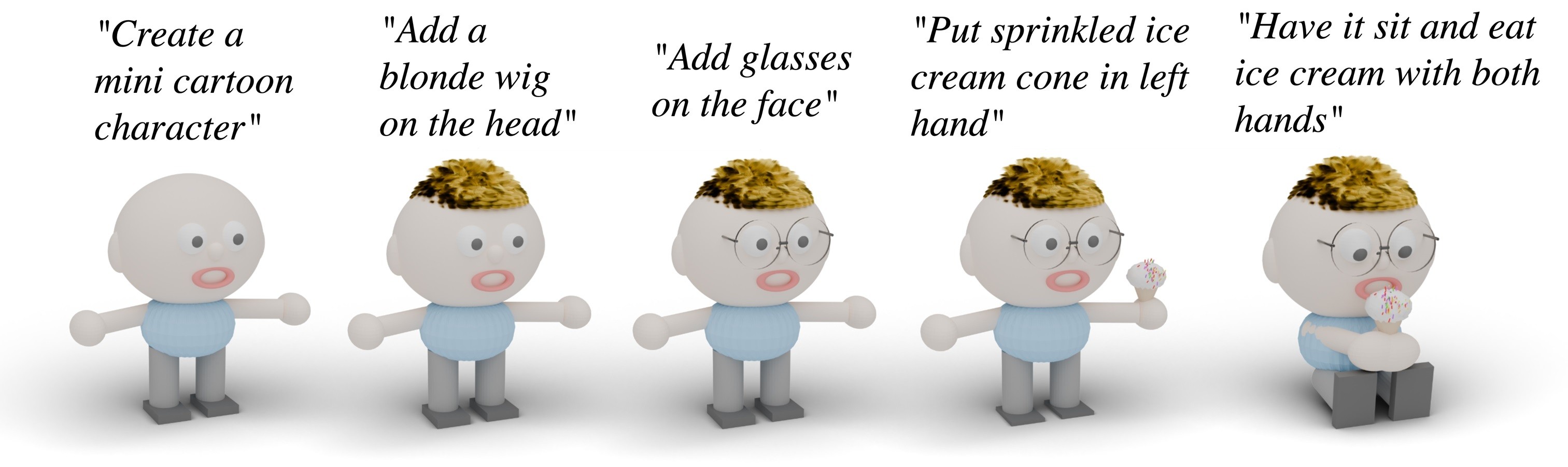}
    \caption{\textbf{Iterative creation.} \ourmethod{} enables multiple successive edits of the same 3D asset. The modifications are faithful to the user's instructions, editing only the specified element while preserving the character’s identity.} 
    \label{fig:humanoid}
\end{figure}

A notable feature of our system is that it enables user-driven, iterative, co-creative 3D modeling.
Instead of generating a 3D shape in a single step, our system optionally enables users to iteratively refine the output through additional 
\rh{
\textit{follow-up prompts} (see \cref{fig:humanoid}). This paradigm contrasts sharply with typical text-input systems, which often require extensive prompt engineering (regenerating the entire result from scratch with each revision). In \ourmethod{}, follow-up prompts target only the relevant part of the shape, preserving the rest of the asset and avoiding the cost of full regeneration. This is possible because we operate on modular, human-readable code and update the existing script rather than rewriting it, enabling an ongoing, collaborative feedback loop between the user and the system.
}

Users can modify the code underlying a system-generated 3D asset, as illustrated in \cref{fig:teaser}, right. Our system writes interpretable, modular code with tunable parameters (\eg knobs and sliders), enabling users to make fine-grained adjustments to the generated shapes with precise control. For example, a user can easily adjust procedural material parameters such as albedo colors, texture, and patterns, or lengths and heights for parametric surfaces in~\cref{fig:intrepretable-slider}. 

A crucial design choice we make is to use retrieval augmented generation (RAG)~\cite{lewis2020RAG} to inject information about writing correct, efficient code for building high-quality 3D shapes. 
Specifically, we build a database containing Blender documentation, referred to as BlenderRAG. BlenderRAG significantly enhances the capabilities of our system without compromising the ability to produce executable code: it enables the LLMs to generate more sophisticated and expressive code by retrieving concrete examples and documentation for advanced operations. 
\begin{figure*}[!t]
    \centering
    \includegraphics[width=\textwidth]{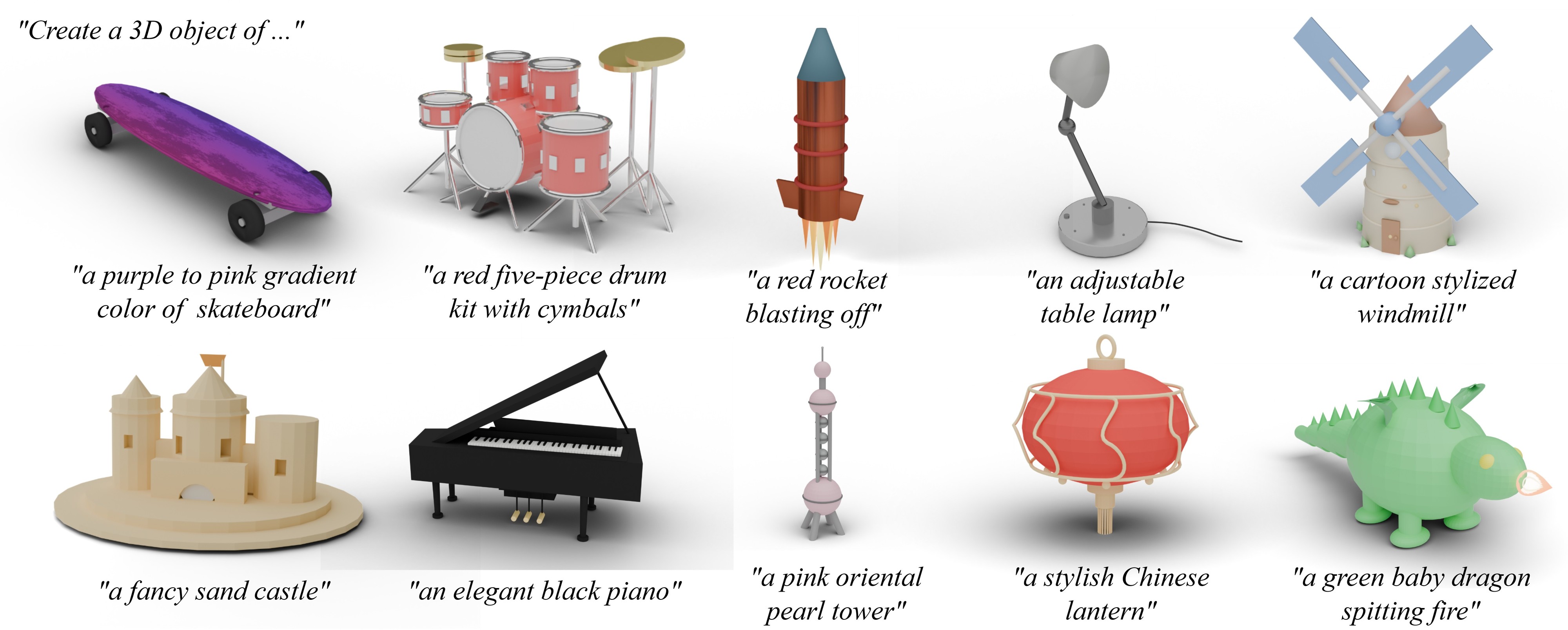}
    \caption{\textbf{Gallery of results.}
    \ourmethod{} is capable of diverse shape generation. The results showcase detailed parts (\eg tassels on the bottom of the lantern, architectural features of the windmill) in intricate arrangements (\eg grouping and spacing of keys on the piano), and rich appearance (\eg gradient color on the skateboard) and material properties (\eg the glossy lamp base). A notable feature of our approach is that each mesh is generated through interpretable, editable Blender code.} 
    \label{fig:gallery}
\end{figure*}

We design a system for generating high-quality 3D assets in three phases, each leveraging multiple specialized agents working in collaboration. In the initial creation phase, we devise a \textit{planner agent} to decompose the user prompt into manageable subtasks, enabling the coding agent to produce a preliminary asset that can be effectively refined in later phases. Next, the auto-refinement phase incorporates visual awareness into the loop. In this phase, a critic agent analyzes renders of the asset, identifies discrepancies with the prompt, and feeds targeted improvement suggestions back to the coding agent, while a verification agent ensures that these fixes are correctly applied. Finally, the user-guided refinement phase allows for integrating direct human feedback, enabling iterative, targeted edits. At the core of \ourmethod{} is representing 3D assets as modular \emph{code} that is shared across all agents, enabling each step of the process to make precise localized code edits.

We present a multi-agent framework that takes text as input and coordinates specialized agents to generate Blender code for creating and editing 3D geometry and appearance attributes. Unlike prior Blender code-writing LLM approaches, which are limited to specific subtasks or constrained procedural programs, our method produces open-vocabulary assets with detailed geometry and rich appearance. With high-level code as a 3D representation, our pipeline is natively a loop of iterative refinement and co-creation: agents perform automatic code and visual self-critique, and users can provide continuous high-level feedback. Further editing avenues are enabled by the clear code and the parameters transparent in the generated Blender nodes and structures.

\section{Related work}
\label{sec:relatedwork}

\noindent \textbf{Representation-Centric Generative 3D Modeling.}
Approaches in generative 3D modeling within graphics and vision commonly aim to design representations of geometry and appearance for use with generative deep learning architectures. Powerful implicit representations such as Neural Radiance Fields~\cite{mildenhall2021nerf} underlie many generative methods such as DreamFusion ~\cite{poole2022dreamfusion} and others~\cite{kosiorek2021nerfvae, wang2022rodin, wang2021clipnerf,ni2025kiss3dgen}. Generative methods have also used the Gaussian splatting representation~\cite{kerbl2023gsplat,chen2023textto3d_gaussiansplatting, yi2023gaussiandreamer, vilesov2023cg3d, lin2024dreampolisher, hu2024turbo3d, wu2025textsplat}. 
Methods also exist for learning latent spaces that are decoded into implicits \cite{zhang20233dshape2vecset,jun2023shape_e,nam2022_3dldm,lan2024ln3diff}.

While these implicit representations excel at high-fidelity visuals, they often lack control, interoperability with graphics pipelines, and interpretability for downstream operations.
As such, a large body of prior work, building on classical geometry processing, explores and operates on explicit shape representations such as point clouds, voxels, and meshes~\cite{chen2021decor_gan,li2021spgan,ibing2021gridimplicit,zheng2022sdfstylegan,zeng2022lion}.
Such results are often directly compatible with the graphics pipeline and human workflows. Notable methods such as PolyGen~\cite{nash2020polygen}, MeshGPT~\cite{siddiqui2024meshgpt}, MeshAnything~\cite{chen2024meshanthing}, and others~\cite{hao2024meshtron,wang2025nautilus,chen2024meshanythingv2artistcreatedmesh} use sequential mesh encodings and autoregressive models to directly predict vertices and faces, aiming to replicate artist-created triangulations.

Other methods approach the same aim via denoising diffusion on triangles~\cite{alliegro2023polydiff,liu2023meshdiffusion,li2025spars3d} or atlases~\cite{yan2024object64}. For the task of editing (rather than generating from scratch), existing methods perform edits on top of input meshes via neural optimization or multiview reconstruction from 2D diffusion priors~\cite{decatur2024paintbrush,dinh2025geometryinstyle,barda2024magicclay,barda2025instant3dit}.

\begin{figure}[!t]
    \centering
    \includegraphics[width=\linewidth]{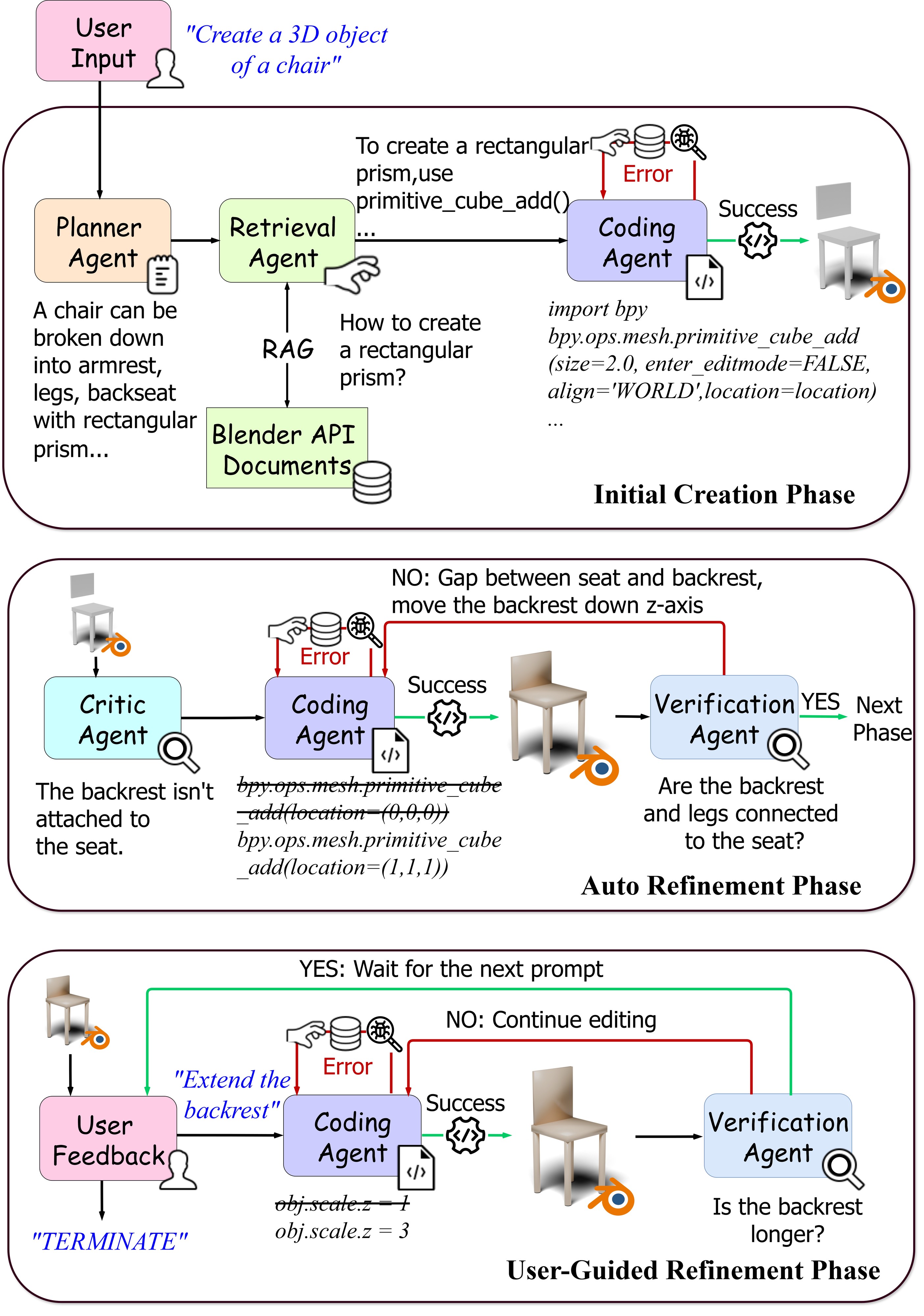}
    \caption{\textbf{Method overview.} Our system uses a team of LLM-powered agents to write Blender code that generates a 3D asset according to a text prompt and additional user instructions. Our method includes three phases: initial creation, automatic refinement, and user-guided refinement. The first phase creates an initial shape, where implausible configurations, like a disconnected backrest, are \textit{automatically} corrected by the second phase. After, our system can accept additional edit instructions from the user, allowing for \textit{interactive} and \textit{iterative} 3D asset generation.}
    \label{fig:overview}
\end{figure}

All of these methods, proposing their own bespoke representations or translations between representations, typically require either training a model from a large collection of 3D shapes or expensive optimizations based on pretrained 2D image models. One of our key contributions is to leverage pretrained LLMs and their general knowledge, aided by RAG, to directly generate explicit 3D shapes in Blender by writing Python scripts, eliminating the need for any additional training \emph{or} neural optimization.

Additionally, many scriptable operations in Blender are high-level, modular, and human-friendly (\eg parametric curves and B-meshes, geometry modifiers, material shaders). This makes the generated code fit well into artist workflows for refining and adjusting 3D objects and are generally amenable to tools in existing graphics pipelines. The geometry and texture of the 3D object can be easily modified by tweaking the parameters, offering more control over the mesh.  

\begin{figure}[!t]
    \centering
    \includegraphics[width=\columnwidth]{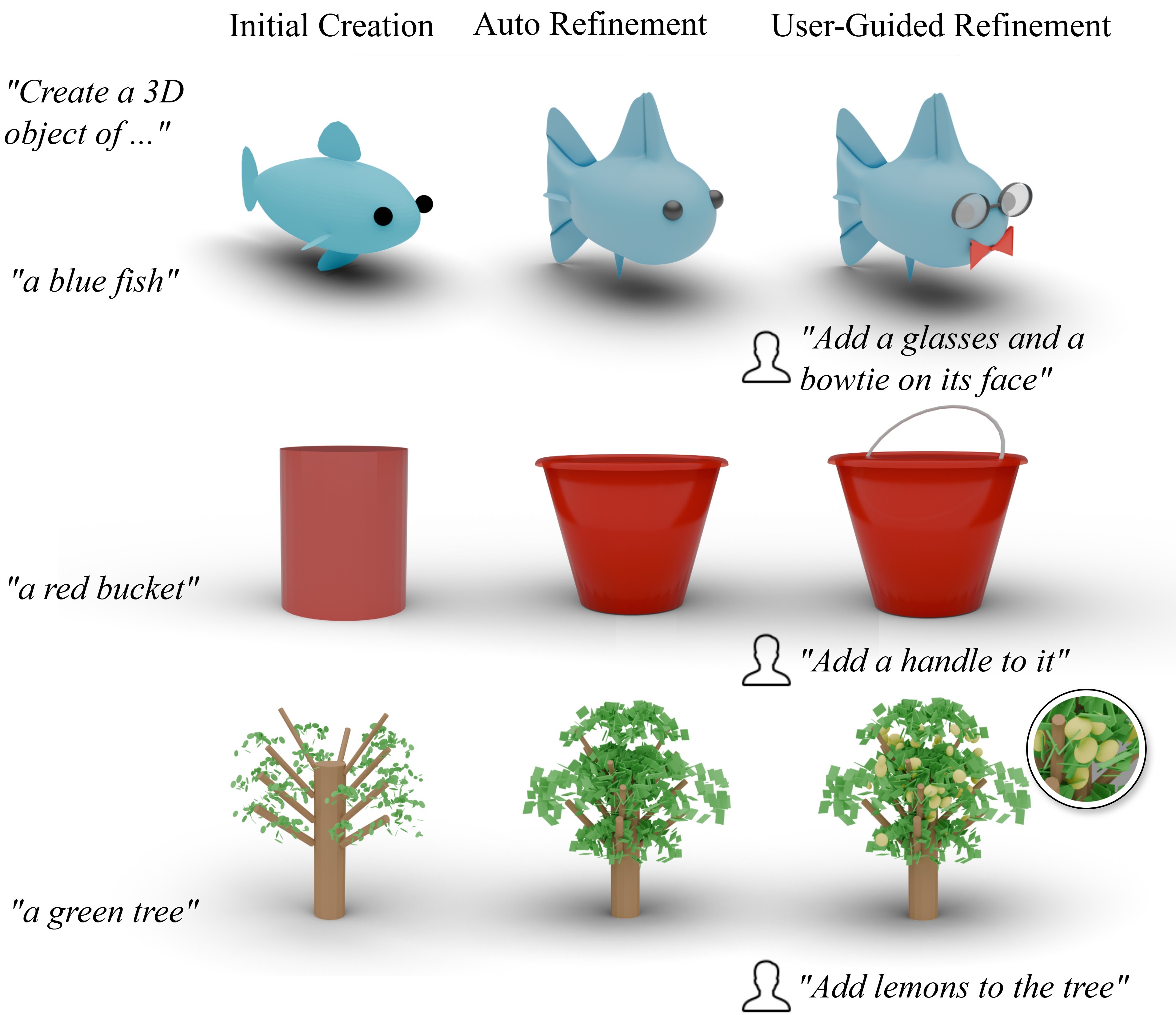}
    \caption{\textbf{Intermediate results for each phase in \ourmethod{}.} Our system enables generating and refining the mesh iteratively. First, the initial creation phase produces a preliminary version of the mesh from the input text (left). Second, the mesh is automatically improved and enhanced through multi-agent visual feedback (middle). Third, the mesh may be refined further based on additional user text input, enabling edits beyond the initial text prompt (right).}
    \label{fig:iterative}
\end{figure}

\begin{figure*}[!t]
    \centering
    \includegraphics[width=\textwidth]{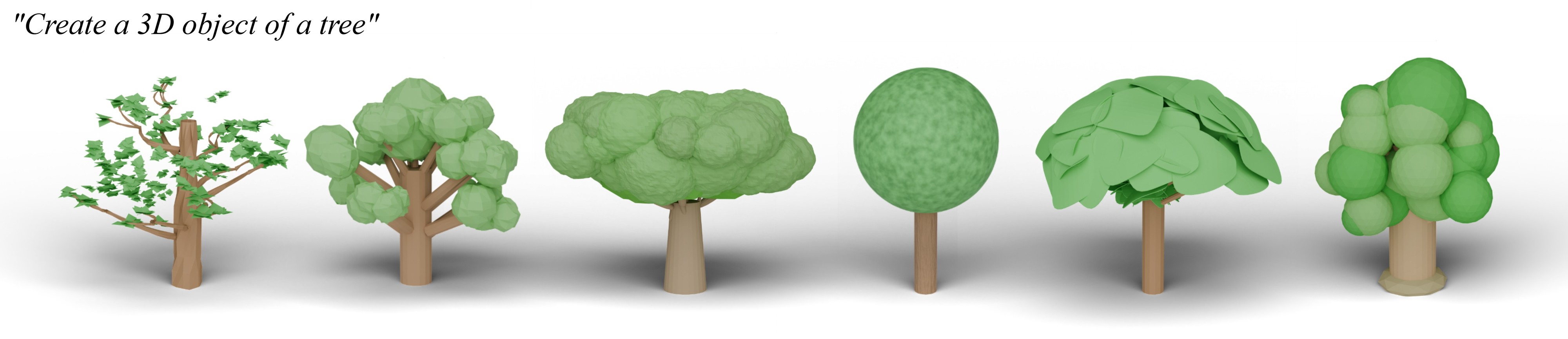}
    \caption{\textbf{Variations.} Due to the inherent stochasticity of language model sampling during inference, identical inputs may yield diverse outputs. This variability also reflects different plausible interpretations of the prompt, often resulting in distinct yet high-quality mesh generations.}
    \label{fig:variation}
\end{figure*}

\noindent \textbf{Program-Based Generation and Editing.}
In contrast to representations that treat shape data monolithically or as continuous functions, human practitioners approach 3D shape modeling with high-level modularity, hierarchy, and abstraction.
Inspired by this, several studies have proposed \emph{domain-specific languages} and \emph{procedural frameworks} for describing 3D objects using primitives and composition procedures. These languages or frameworks have parameters and programs that may be learned by deep generative models. Prior work includes: GeoCode \cite{pearl2025geocode}, parameterized geometry node programs; PyTorchGeoNodes \cite{stekovic2025pytorchgeonodes}, differentiable and optimizable representations of Blender geometry node programs; ShapeAssembly \cite{jones2020shapeasm}, a domain-specific language (DSL) for hierarchically describing shapes, as well as associated methods \cite{jones2021shapemod,jones2023shapecoder,ganeshan2024parsel,jones2024visualprogramedit} that train models, discover decompositions, and reconstruct shapes in this high-level language. 

Being high-level abstractions, these representations can lead to more efficient generative models with results that are more interpretable to artists and downstream tasks. However, such bespoke representations are often category-specific, or make other restrictive assumptions (\eg on the type of geometric primitive) to create shapes.

\smallskip
\noindent \textbf{LLMs for Shape Generation and Editing.} The success of LLMs to break new ground in NLP has inspired research in visual domains to leverage LLMs for a broad variety of tasks~\cite{jiaxi2023gpt4motion,zhangblendergym2024,vinker2025sketchagent}. Recent work has begun to use LLMs to generate meshes directly, such as through mesh token prediction~\cite{taomeshllm2024,wang2023llamamesh,lonar2025treemeshgpt}. However, mesh elements are a low-level representation that lacks high-level shared structure across different examples (as opposed to code which does, see \cref{fig:shared_code_structure}), and does not support high-level interpretable parameter changes.

\begin{figure}[!b]
    \centering
    \includegraphics[width=\columnwidth]{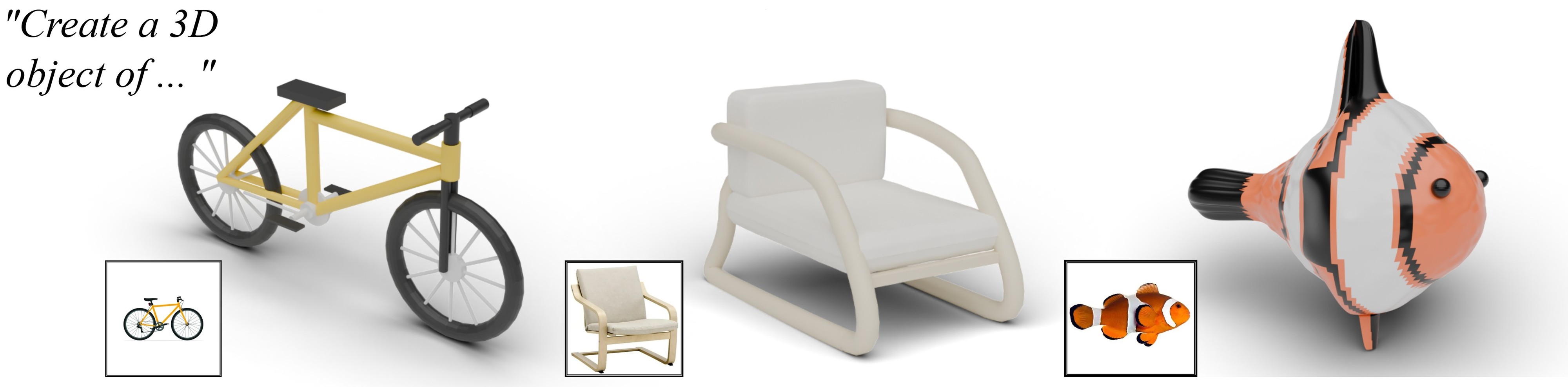}
    \caption{\textbf{Image input.} \ourmethod{} supports images (along with a corresponding text instruction) as an initial multi-modal input, enabling the system to capture and translate visual details into 3D assets.}
    \label{fig:image-prompt}
\end{figure}

To address this, several works have used LLMs to generate Blender Python code. L3GO~\cite{yutaro2024l3go} generates simple primitive shapes without appearance attributes in Blender. BlenderAlchemy~\cite{ian2024blenderalchemy} generates materials in Blender for existing geometry. 3D-GPT~\cite{chunyi20233dgpt} generates parameters for Infinigen, a preexisting procedural generator with predefined scenes (primarily natural environments). SceneCraft~\cite{ziniu2024scenecraft} retrieves 3D assets and uses an LLM to organize them in a coherent spatial scene layout. Different than the aforementioned approaches, our system generates complete 3D assets, including detailed geometry with appearance attributes, without any predefined class constraints.

\begin{figure}[!t]
    \centering 
    \includegraphics[width=\columnwidth]{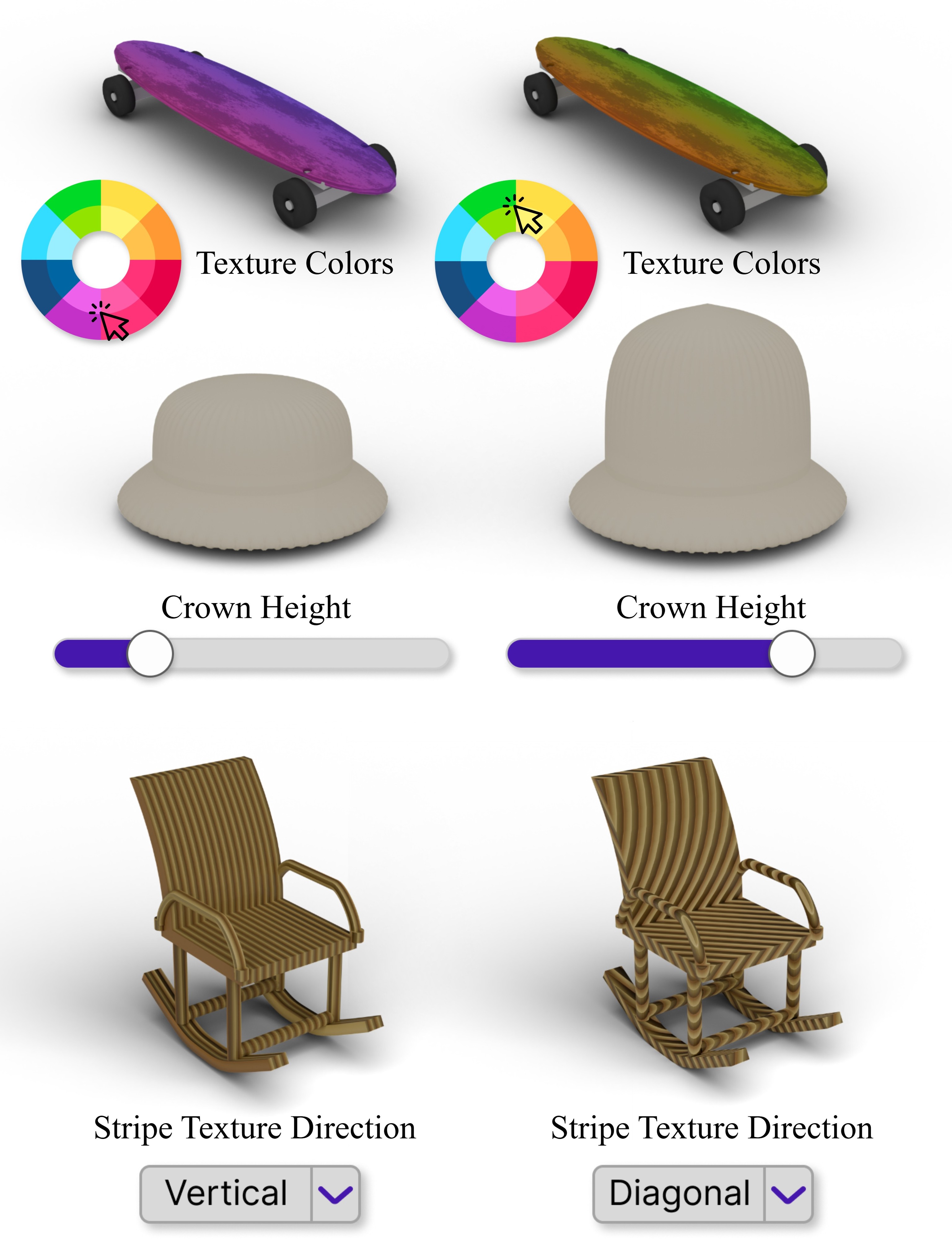}
    \caption{\textbf{Interpretable edit parameters.} By generating shapes through Blender code, LL3M enables intuitive user edits via interpretable parameters, such as those exposed by geometry and shader nodes.
    For example, when generating a material, our system creates a full set of Blender shader nodes. Users can then easily adjust visual attributes, such as tuning the color or stripe pattern, directly in Blender, to achieve the desired output.}
    \label{fig:intrepretable-slider}
\end{figure}

The most relevant method to ours is BlenderMCP~\cite{blendermcp} which can also generate complete 3D assets by writing code. BlenderMCP uses a single LLM (\eg Claude) calling Blender functions via the Model Context Protocol~\cite{mcp2024website}. We compare against this baseline in \cref{fig:comparsion} and observe that our multi-agent system provides superior quality assets with more detailed geometry. In addition, a notable and unique feature of our approach is the many options for users to iterate on and refine the generated 3D asset (see \cref{fig:humanoid}, \cref{fig:intrepretable-slider}).

\section{Method}
\label{sec:method}

\begin{figure}[!b]
    \centering
    \includegraphics[width=\columnwidth]{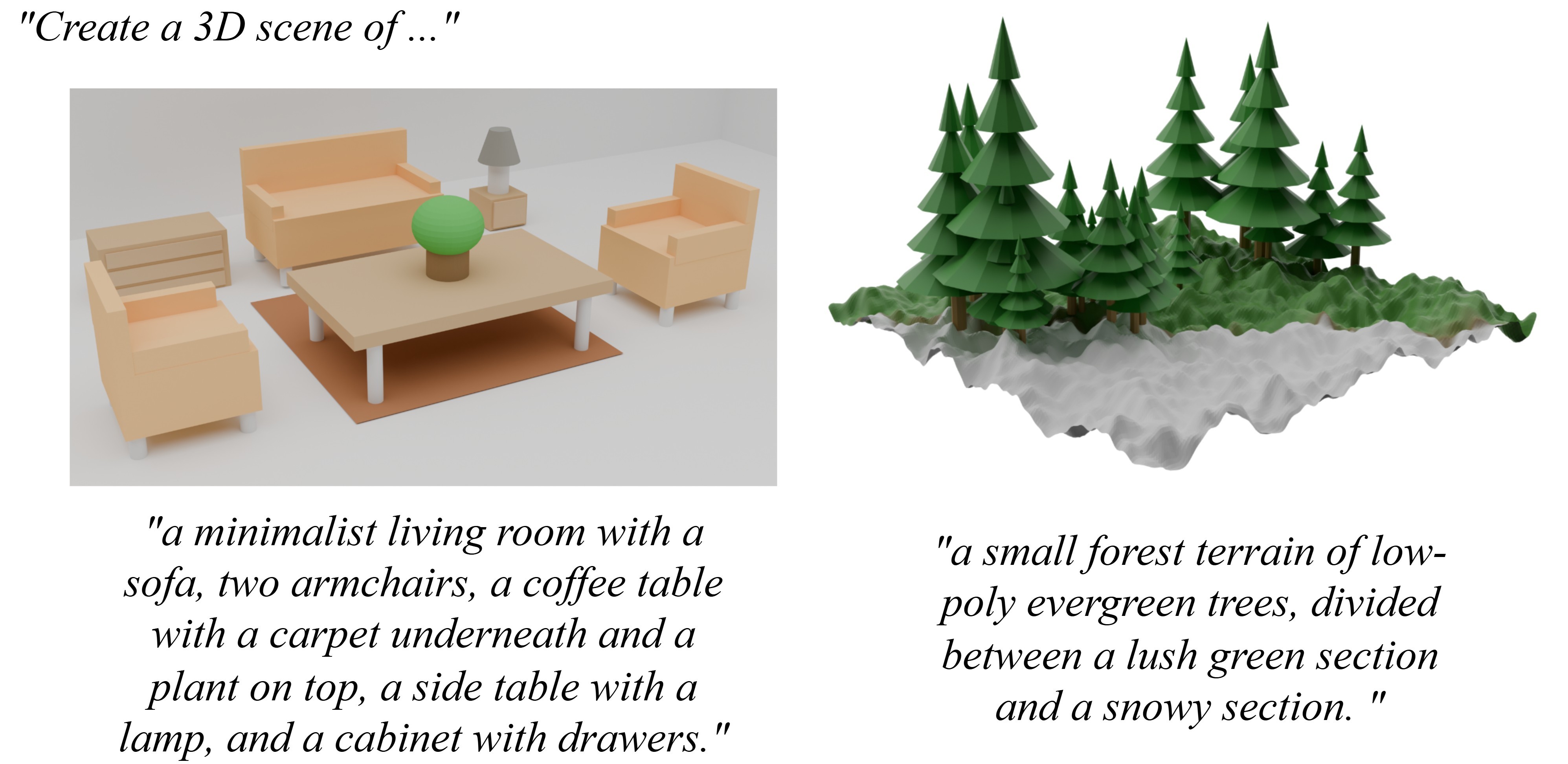}
    \caption{\textbf{Scene generation.} \ourmethod{} is capable of generating multiple objects and arranging them with appropriate spatial relationships within a single scene. Our system achieves this task using complex operations such as instancing and parenting relationships to build the scene hierarchy.}
    \label{fig:scene}
\end{figure}

Given an input text prompt, our goal is to generate a 3D asset in Blender that aligns with text specifications and enables further iterative editing. To achieve this task, we propose \ourmethod{}, a multi-agent framework, where agents collaborate to write Blender code to incrementally construct, refine, and edit the desired 3D asset.

An agent is an entity, powered by a Large Language Model (LLM), that participates in a multi-agent conversation to solve the 3D modeling task. Each agent has a specific role and a set of instructions to follow and/or tools to use (where a \emph{tool} is some external executable function that the LLM can issue a call to, possibly receiving its results.) Our pipeline consists of six agents: planner, retrieval, coding, critic, verification, and user feedback. The role definition of each agent and its instructions and tools are elaborated in the supplementary material. 

The \textit{planner agent} is in charge of breaking down the user's initial prompt into 3D modeling subtasks and delegating them to the retrieval and coding agents. The \textit{retrieval agent} retrieves relevant Blender documentation and summarizes it. Its primary purpose is to provide high-quality, complex code usage examples to the coding agent and help it debug coding errors. The \textit{coding agent} writes Blender code and executes it. The \textit{critic agent} looks for visual problems of the generated asset and proposes fixes. The \textit{verification agent} checks whether the proposed changes by the critic agent were correctly implemented by the coding agent. Lastly, the \textit{user agent} receives and processes additional user inputs for further editing of the 3D asset.

The agents are orchestrated by an external controller that handles their communication and order. This allows shared contexts between all agents, meaning that one agent can see the output of every other agent, and the conversation history of the agents is shared across the 3D asset generation. For more implementation details about agent management, please see the supplementary. 

Our approach consists of a three-phase process: initial creation, auto-refinement, and user-guided refinement, as illustrated in \cref{fig:overview} and demonstrated in \cref{fig:iterative}. In the initial creation phase, the system generates the core geometric and appearance components of the 3D asset. Then, in the auto-refinement phase, the visual critic agent gives feedback on the asset quality, style, materials, and connection between parts. Finally, the user-guided refinement phase allows users to iteratively refine the 3D model beyond the initial text input, enabling greater customization to match their preferences. We elaborate on each phase in \cref{subsec:initial_creation_phase,subsec:auto_refinement_phase,subsec:user_guided_refinement_phase}.

\subsection{Initial Creation Phase}
\label{subsec:initial_creation_phase}
The first stage is the \textit{initial creation phase}, which consists of the \textit{planner agent}, the \textit{retrieval agent}, and the \textit{coding agent}. The workflow begins with the user providing an input text prompt denoted as task $T$. The planner agent $\Acal_p$ takes as input $T$ and decomposes it into $n$ specific subtasks
\begin{equation} \label{eq:planar_agent}
\{T_1,T_2,...,T_n\} = \Acal_p(T).
\end{equation}

\noindent As an example, the generation of a chair ($T$) may be broken down into the generation of legs ($T_1$), a backrest ($T_2$), and a seat ($T_3$). The retrieval agent $\Acal_r$ then takes a subtask $T_i$ as input to call a tool to query the BlenderRAG database, which retrieves the relevant Blender documentation and summarizes it as $R_i$:
\begin{equation} \label{eq:retrieval_agent}
R_i = \Acal_r(T_i),\;\;i = 1, ..., n.
\end{equation}
The coding agent $\Acal_c$ considers the subtask $T_i$ and the Blender documentation summary as input to write and execute the code in Blender. The code for the previous subtasks is also included as input from the shared context. Its complete code output is:
\begin{equation} \label{eq:coding_agent}
C_i = \Acal_c(T_i, R_i, \{C_1,...,C_{i-1}\}),\;\;i = 1, ..., n.
\end{equation}
where $C_i$ is the code for a 3D asset associated with the corresponding subtask $T_i$. \noindent If an execution error occurs, the retrieval agent takes the error message, denoted as $E_i$, and calls the tool to query BlenderRAG for the solution (\eg, error: \textit{key ``Specular" not found}, solution: \textit{the key has been renamed to ``Specular IOR Level" for Blender 4.X}). This process of error correction by the retrieval and coding agents repeats until the code execution does not yield an error. The retrieval agent then takes the next subtask as input and repeats the process above until all subtasks are completed. This concludes the initial creation phase, and the system moves on to the next phase. The output of this phase is an error-free Blender code $C_n$ that creates a 3D asset that roughly corresponds to the user's initial input.

\subsection{Automatic Refinement Phase}
\label{subsec:auto_refinement_phase}

The second phase is the auto-refinement phase, consisting of the \textit{critic agent}, the \textit{coding agent}, and the \textit{verification agent}. Although the initial creation phase produces executable Blender code that creates a 3D asset, the asset often contains flaws such as disconnected geometry or visual attributes that fail to align with the input prompt. To address this, the auto-refinement phase is introduced to automatically improve the asset quality by incorporating visual awareness into the process.

In this phase, the critic agent receives the task $T$ and renders $m$ images $\{I^q_1,...,I^q_m\}$ of the current 3D asset in Blender using adaptive camera distance based on the object’s bounding box. Then, the critic agent calls a tool to pass the rendered images and a pre-defined prompt $P$ into an external vision-language model (VLM) to analyze and identify visual issues and propose ways to improve them (\eg, problem: \textit{The legs aren't attached to the seat}, solution: \textit{Move the legs up the z-axis}). The pre-defined prompt $P$ asks the VLM to check if the renders match the task $T$. The output of the critic agent $\Acal_q$ is:

\begin{equation} \label{eq:critic_agent}
\{Q_1,Q_2,...,Q_n\} = \Acal_q(\{I^q_1,...,I^q_m\}, P).
\end{equation}

\noindent where $Q_i$ is the critique for subtask $T_i$, including the visual issues and suggested fixes. If the critic identifies problems with the current 3D asset, its feedback $\{Q_1,...,Q_n\}$ and the current script $C$ are passed to the coding agent for improvement. 

Crucially, providing the existing Blender code $C$ to the coding agent enables modifying certain components of the code (and thus the shape) as opposed to rewriting the code from scratch. Without providing the existing Blender code to the coding agent, the coding agent will rewrite the code to produce a different 3D asset, as shown in \cref{fig:context}. After the code is refined based on the proposed improvements, the coding agent executes the modified Blender code. Similar to the initial creation phase, the retrieval agent will be invoked if a compilation or execution error occurs. 

However, it may not always resolve all the issues in one attempt. For example, in \cref{fig:verification} of \cref{subsec:multi_agent_coordination}, the coding agent addresses critiques 2 and 3 for the fire hydrant mesh, but critique 1 was not fixed. To address this, we introduce a verification agent, also driven by a VLM. This agent calls a tool to render the current 3D asset $\{I^v_1,...,I^v_m\}$. It then passes the most recent renders ($\{I^v_1,...,I^v_m\}$), the previous set of renders from the critic agent ($\{I^q_1,...,I^q_m\}$), and the critiques ($\{Q_1,...,Q_n\}$ into an external VLM and verifies whether the proposed solutions of each critique have been successfully applied (\eg \cref{fig:verification}, problem: \textit{Partially, the cap isn't attached to the body}, solution: \textit{Move the cap further down along z-axis}). The output of the verification agent $\Acal_v$ are verification instructions for the coding agent:
\begin{equation} \label{eq:verification_agent}
\begin{aligned}
\{V_1,...&,V_n\} = \\
&\Acal_v(\{I^v_1,...,I^v_m\}, \{I^q_1,...,I^q_m\}, \{Q_1,...,Q_n\}).
\end{aligned}
\end{equation}

\begin{figure}
    \centering 
    \includegraphics[width=\linewidth]{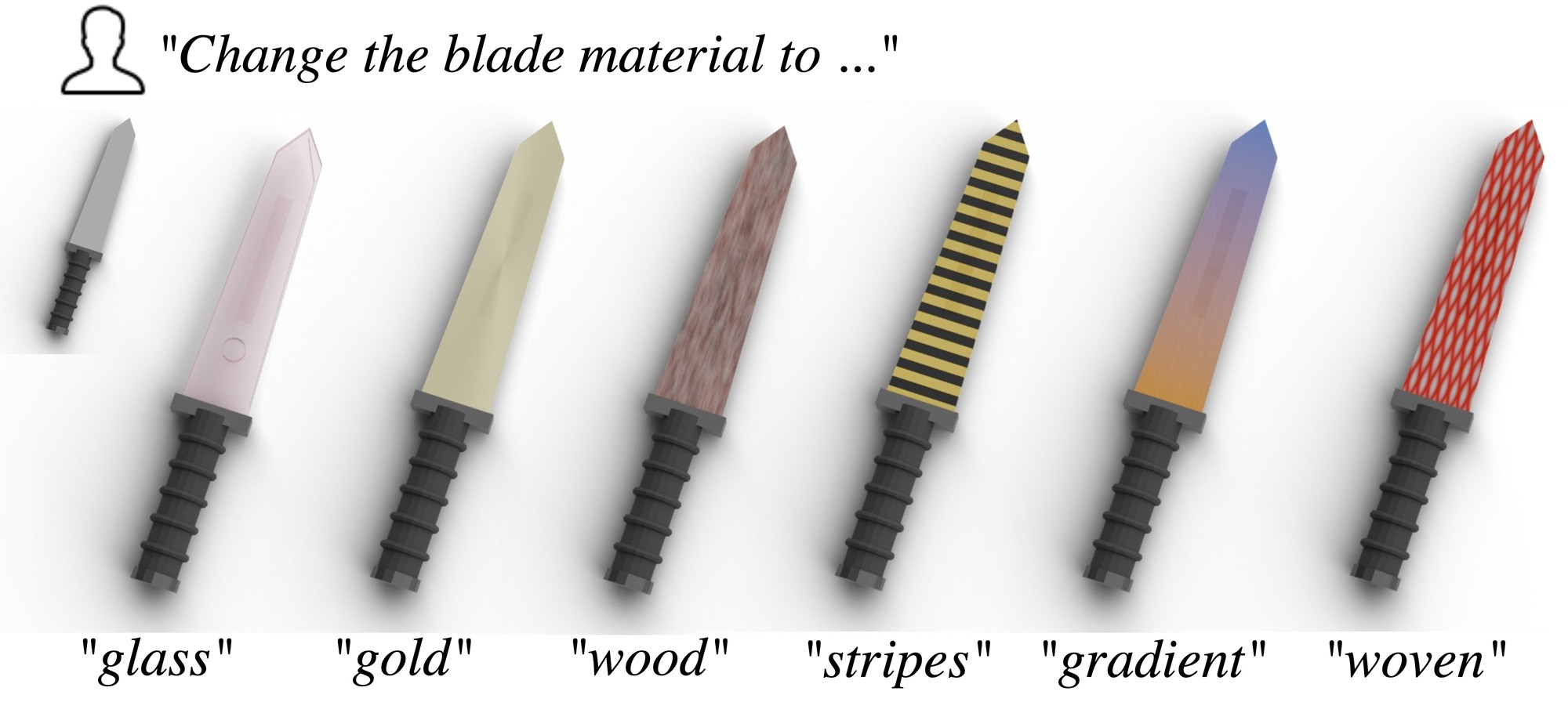}
    \caption{\textbf{Material editing.} Given an initial mesh produced by our system, our system is capable of editing the materials on a specific part of the mesh (the blade of the knife), by creating comprehensive procedural materials via shader nodes.}
    \label{fig:same-in-diff-mat}
\end{figure}

\begin{figure*}[!t]
    \centering
    \includegraphics[width=\textwidth]{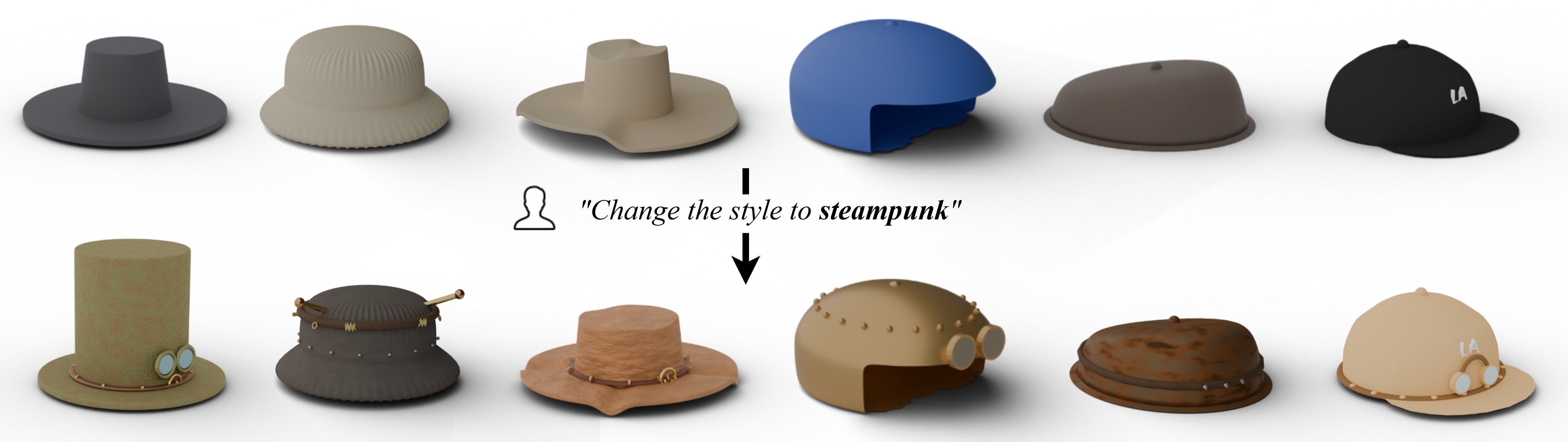}
    \caption{\textbf{Consistent stylization.} Starting from different initial meshes produced by \ourmethod{} and the same refinement prompt ``change the style to steampunk,” \ourmethod{} successfully interprets and applies the same style concept to each hat. Each stylized mesh produces distinct variations, including both geometric modifications and appearance changes.}
    \label{fig:diff-in-same-out}
\end{figure*}

If the 3D model is not approved by the verification agent, its output and the current shape generation code $C$ are passed to the coding agent for refinement. Otherwise, the process continues to the next phase. The output of this phase is Blender code of a 3D asset that closely aligns with the user's initial input.

\subsection{User Guided Refinement Phase}
\label{subsec:user_guided_refinement_phase}

The third phase is the user refinement phase, which consists of the \textit{user agent}, the \textit{coding agent}, and the \textit{verification agent}. This phase allows users to provide additional instructions to modify the existing mesh and extend beyond the initial text input, as shown in \cref{fig:same-in-diff-out,fig:same-in-diff-mat,fig:diff-in-same-out}.

While the mesh is automatically generated at the end of phase two, users may wish to additionally modify it to better suit their preferences. For example, the user may want to make geometric changes, like sculpting a face in a wizard's hat in \cref{fig:teaser}, or stylistic changes, like adding a steampunk-style twist in \cref{fig:diff-in-same-out}.

Since Blender's code is highly modular and reusable, changing the geometry or materials can be made by adding a few lines of code, see \cref{fig:teaser}, or adjusting features of the 3D asset in Blender through shader nodes, as shown in \cref{fig:intrepretable-slider}. In other words, the user can either provide additional text prompts via a user agent in this phase or \textit{directly} modify the code in Blender.

If the user decides to provide additional text prompts, then they can simply provide high-level instructions like \textit{``add lemons to the tree''}, as shown in \cref{fig:iterative}). We denote this additional text prompt as $T'$. The coding agent takes $T'$, updates the current Blender script, and executes it. The same verification agent from the auto-refinement phase is then invoked to confirm if the changes from the additional text prompt are properly implemented. If the changes are successfully applied, the system returns to the user agent, awaiting the next text prompt. This cycle continues until the user explicitly provides a keyword to terminate the process. Then, the workflow stops and the user is given a complete, tailored 3D asset in Blender that aligns with both the initial input prompt and any edits they requested after the fact, along with the code $C$ for generating the asset. For more complex meshes, multiple prompts may be required to adhere to what the user wants. An example of \cref{fig:humanoid} is in \cref{sec:user_guidance} in the supplementary.

\section{Experiments}
\label{sec:experiments}

In this section, we first detail some concrete implementation choices in \cref{subsec:impl}. We showcase the properties of our method in \cref{subsec:properties_of_ourmethod}, including its generality, fidelity, and interpretability. Then, we demonstrate the high flexibility during the user-guided refinement phase in \cref{subsec:user_guided_refinement}. Finally, \cref{subsec:evaluation} presents qualitative comparisons, quantitative evaluations, and ablation experiments. 

\subsection{Implementation details}
\label{subsec:impl}
\ourmethod{} is implemented in the AutoGen framework~\cite{wu2024autogen}, which enables a coordinated and collaborative task-solving process across multiple agents. BlenderRAG is implemented with RAGFlow~\cite{ragflow}, a tool for building RAG databases and interfacing them with LLMs. 

We used Blender 4.4~\cite{blendersoftware} for our experiments. To construct BlenderRAG, we converted 1,729 official Blender 4.4 documentation HTML files into PDFs and injected them into a knowledge base in RAGFlow, enabling the retrieval agent to access version-specific knowledge for the Blender Python module. 
We note that upon future releases of Blender, BlenderRAG can easily be updated with the new documentation. In this way, \ourmethod{} can make full use of the latest (or a specific) Blender version, regardless of the coding agent's knowledge of Blender versions from its pretraining.
For the critic and verification agents, we render $m=5$ views. Please refer to the supplementary material for examples of the renders. 

Each agent is powered by an LLM. We employ GPT-4o~\cite{openaigpt4o} for the planner and retrieval agents for its best general planning ability, Claude 3.7 Sonnet~\cite{claudesonnet} for the coding agent for its high performance in coding benchmarks \cite{livebench2025}, and Gemini 2.0 flash~\cite{gemini2flash} for the critic and verification agent, given its visual analysis capabilities.  Additional details of each agent are given in \cref{sec:agent_implementation_details}.

\begin{figure}
    \centering
    \includegraphics[width=\columnwidth]{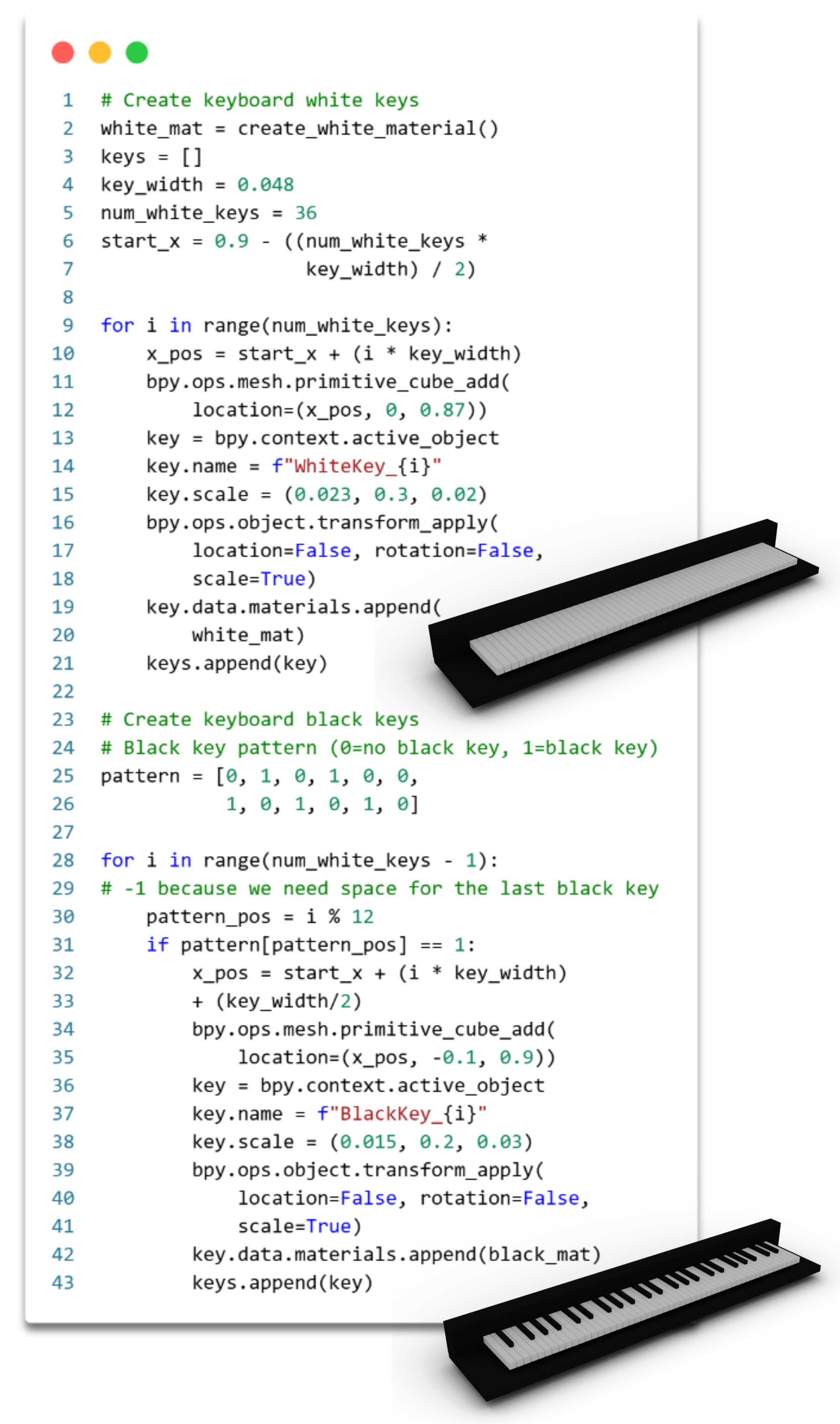}
    \caption{\textbf{Interpretable code.} Our method generates Blender code that is easy to understand and follow. The code is well-documented with descriptive comments, clear variable names, and structured logic. This interpretable code makes it easy to potentially change variables (\eg the key width) or even algorithmic logic (\eg the keyboard pattern).} 
    \label{fig:snippet}
\end{figure}

\subsection{Properties of \ourmethod{}}
\label{subsec:properties_of_ourmethod}

\begin{figure}[!t]
    \centering
    \includegraphics[width=\columnwidth]{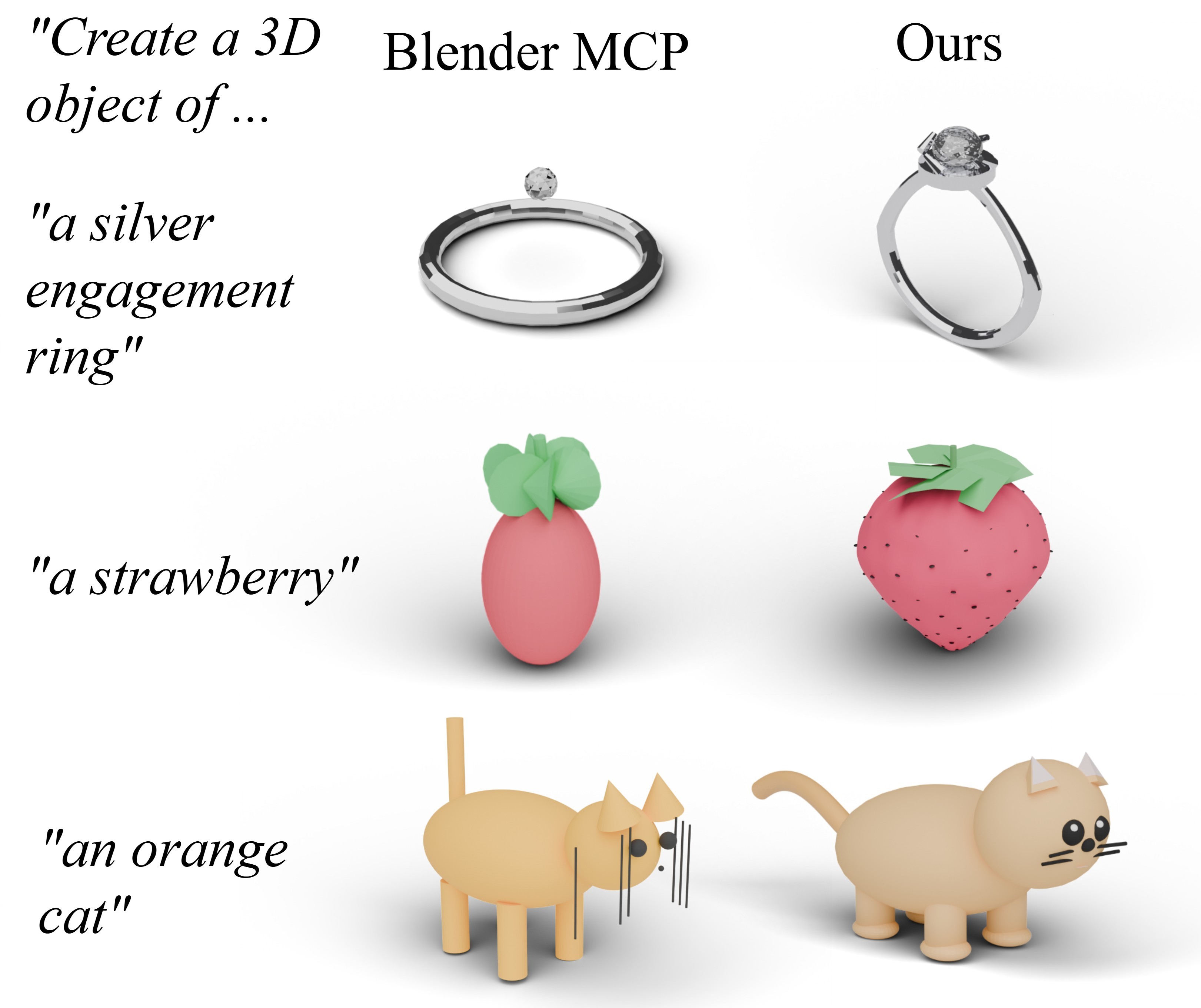}
    \caption{\textbf{Baseline comparison.} We compare the Blender MCP module (left) with our method (right)~\cite{blendermcp}. Our method produces higher-quality meshes with more detailed geometry and closer alignment to the input text prompt compared to the baseline.}
    \label{fig:comparsion}
\end{figure}

\smallskip
\noindent \textbf{Generality and versatility.} Our method is general and flexible. \ourmethod{} is able to create 3D assets from a variety of categories, ranging from vehicles, musical instruments, fruits, manufactured objects, and animals, to indoor and outdoor scenes, as demonstrated in \cref{fig:gallery,fig:image-prompt,fig:comparsion,fig:scene}. \ourmethod{} generates detailed 3D geometry, in addition to diverse properties, including texture, material, part hierarchy, and more. 

\cref{fig:gallery} showcases a variety of generated assets. As seen in the figure, the shapes vary in geometry, texture, and material. For example, the dragon's spikes and the windmill's turbine are composed of sharp elements, while the Chinese lantern and the rocket have curved geometry. Our method also creates intricate structures, like the drum kit's stands and the piano keyboard. In addition to the geometry, the assets may contain complex textures, such as the gradient colored skateboard, or glossy material in the lamp. %

Given the same text input, our method will create plausible variations, as seen in \cref{fig:variation}. For a generic prompt such as \textit{``create a 3D object of a tree''}, without additional specifications, \ourmethod{} generates substantially different trees, with varying geometry for the trunk, branches, and foliage.

\smallskip
\noindent \textbf{Fidelity.} \cref{fig:iterative} presents the intermediate results for each of the three phases in our system.
\ourmethod{} exhibits high fidelity to the user's input text prompt. This is enabled by our auto-refinement phase that improves the plausibility of the asset generated in the initial creation phase, and improves the adherence to the initial prompt.

For the prompt ``a red bucket" in \cref{fig:iterative}, the initial creation phase produces a red diffuse cylinder, but its shape does not yet resemble a bucket. During the auto-refinement phase, the material is updated to resemble reflective plastic, and the bucket’s body is modified to taper inward with a lip at the top, resulting in a more accurate and realistic 3D asset. Similarly, for the prompt ``a tree", more leaves and branches are added to enhance its plausibility. Then, the edits in the user refinement phase accurately follow the user's instructions. Glasses and a bow tie are added to the face at the correct locations, while the rest of the fish remains intact.

\cref{fig:gallery} showcases additional color and texture fidelity examples, like the red rocket and the gradient-colored skateboard. Our system demonstrates contextual understanding, the rocket is described as \textit{``blasting off"}, and the system generates the rocket object and an additional fire blast at its base to reflect the action.

The main focus of our system is generating a single asset using a text prompt as input. However, our system also supports image inputs (\cref{fig:image-prompt}), and can generate multiple objects and coherently place them within a single scene (\cref{fig:scene}). These use cases further demonstrate the fidelity and flexibility of our approach. For example, the 3D bicycle in \cref{fig:image-prompt} accurately reflects both the color and structure shown in the image, capturing fine-grained details such as the pedals and spokes. Likewise, the generated scenes in \cref{fig:scene} contains the various elements described in the elaborated and detailed prompts (sofa, chairs, coffee table for the indoor scene, and trees, lush green and snowy sections for the outdoor scene), and respect their style (minimalist living room and low-poly trees, respectively).

\smallskip \textbf{Structured and interpretable generation.} 
An example of our system-generated code can be found in \cref{fig:snippet}, which is notably easy to interpret. The code includes comments explaining each component’s functionality, uses intuitive variable names, and follows a clear, easy-to-understand logic. 
For example, the piano key generation code uses two \texttt{for} loops: one to create the specified number of white keys (\texttt{num\_white\_keys}) and another to add black keys following a defined pattern (\texttt{pattern}). The code even explains the use of \texttt{-1} in the second loop, which accounts for adding the final single black key, just as in a real-world piano.

This unique property of 3D asset generation via code enables precise local edits and intuitive shape manipulations. Geometry created by our system can be edited by parameter changes in the code (like the fish in \cref{fig:teaser}), or additional elements can be added by incorporating new code without changing the existing code, keeping the remaining of the shape intact (like adding the candle to the lantern in \cref{fig:teaser}). Additional user modifications are detailed in \cref{subsec:user_guided_refinement} and in the supplementary material.

\subsection{User-guided Refinement}
\label{subsec:user_guided_refinement}

\noindent \textbf{Material modification.} In addition to geometric edits (\cref{fig:teaser}), the user-guided refinement phase in our system supports dedicated material changes, demonstrated in \cref{fig:same-in-diff-mat}. Given a 3D knife created by our system, \ourmethod{} can create and apply different materials to create appearance attributes such as texture, color, roughness, and more. Because our code is structured and modular, it supports precise local modifications, such as automatically editing only the blade region, while leaving all other parts of the asset unchanged.

\smallskip
\noindent \textbf{Stylization modification.} Beyond concrete edit requests, our system can also interpret high-level, potentially ambiguous instructions, such as themes or styles, as shown in \cref{fig:diff-in-same-out}). Given different initially styled hats and the same style guidance prompt, our system interprets and adapts the style concept to suit each hat. Each result preserves the hat’s original overall shape and type, while adding materials, colors, surface geometric details, and even extra elements, such as small goggles, to vividly fulfill the \emph{steampunk} style instruction.

\noindent \textbf{Interpretable user edits.} The interpretable code written by \ourmethod{} allows users to \textit{directly} modify it, such as tweaking the parameter values to suit their preferences. The modification can be applied to different aspects of the asset, including geometry, texture, color, and patterns. For example, in \cref{fig:intrepretable-slider}, users can precisely control the crown height of the hat by adjusting the value of this parameter. Additionally, the specific texture color of the skateboard can be selected from a color palette, and the direction of the stripes pattern on the chair is determined by a simple drop-down selection. We note that each edit influences only the intended shape property; texture modifications do not change the geometry and vice versa.

\noindent \textbf{Iterative creation.} \ourmethod{} supports successive modifications of the generated shape, facilitating an iterative creation process, exemplified in \cref{fig:humanoid}. The created character can be modified in different ways. The user can add new attributes, such as the wig, glasses, and ice cream, or change existing attributes, like the character's pose. We highlight that shape manipulations closely follow the user's requests. For example, the wig has the specified color, and the ice cream is added to the correct hand. The system can interpret semantic instructions and apply them with fine-grained semantic edits. Specifically, the \textit{``eating the ice cream''} instructions are correctly applied by bringing the ice cream to the mouth region. The system also adjusts the character’s pupils downward, making it appear as though the character is looking at the ice cream it is eating. This example demonstrates the collaborative generation process offered by \ourmethod{}. 

\subsection{Evaluation}
\label{subsec:evaluation}

\noindent \textbf{Qualitative comparison.} 
Our system differs from existing approaches in that it generates Blender code that generates complete 3D assets, including detailed geometry with appearance attributes, without any predefined class constraints. Among prior approaches, the most relevant comparison we identify is BlenderMCP~\cite{blendermcp}, a method that claims an ability to create arbitrary shapes from text via a single LLM (Claude), in a single conversation context, calling Blender functions. \cref{fig:comparsion} presents the results of BlenderMCP and \ourmethod{}, where we use the same text input for both methods. Our method produces higher-quality 3D shapes with richer details and greater plausibility compared to BlenderMCP. For example, the cat's whiskers are in the correct orientation and properly connected to its face, and the strawberry has seeds.

\begin{figure}[!t]
    \centering
    \includegraphics[width=\columnwidth]{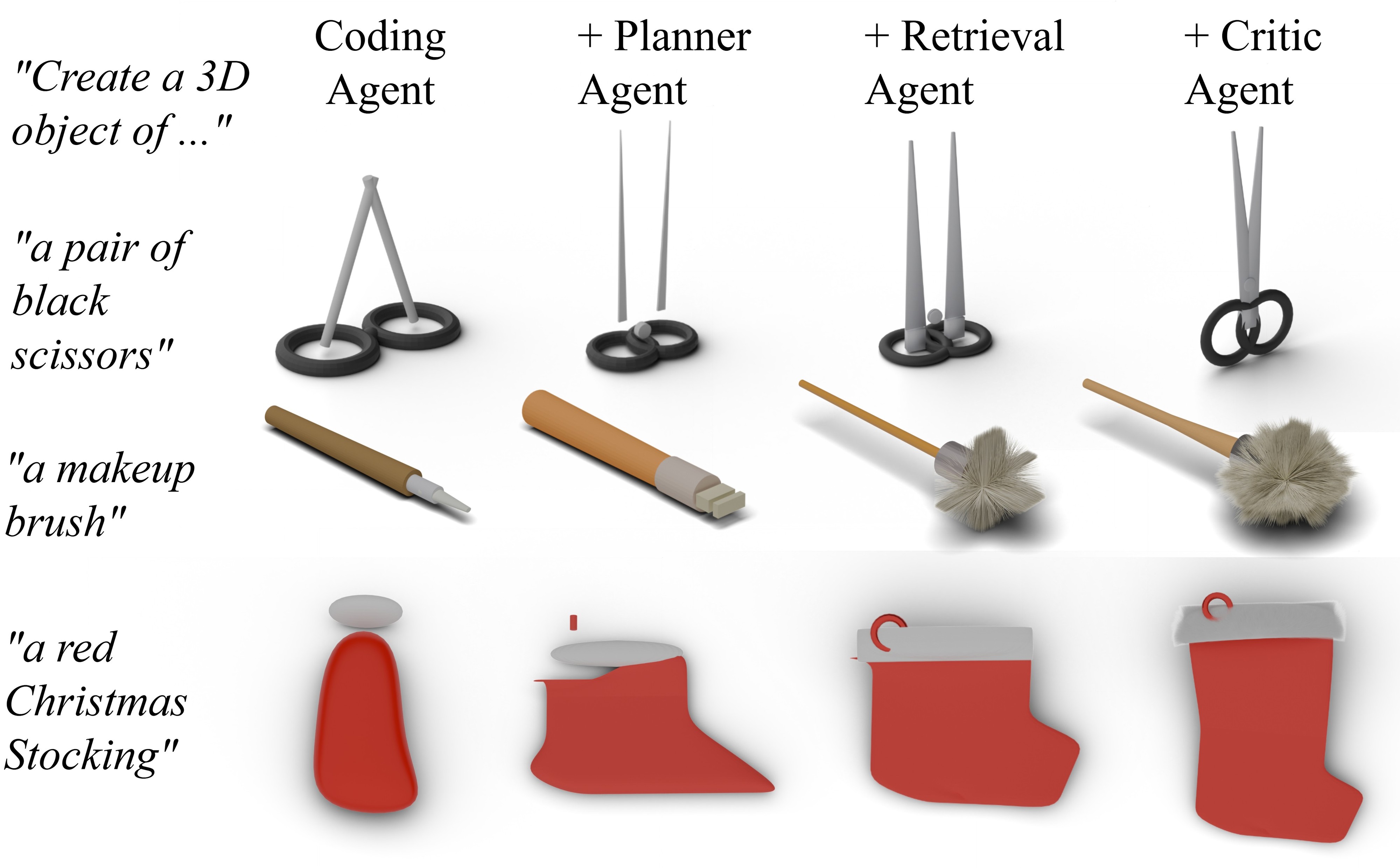}
    \caption{\textbf{Ablation.} Our system consists of several specialized agents who work together to generate the 3D asset. Adding each of the agents assists in further improving the shape's quality and its fidelity to the input text, indicating their importance in the method.}
    \label{fig:ablation-agent}
\end{figure}

We attribute the improved quality of our results to multiple critical design choices in our system, several of which are visually highlighted in \cref{fig:ablation-agent}. We devise multiple agents with specific tasks that collaborate and coordinate together to design sophisticated 3D assets. 

\smallskip \noindent \textbf{Quantitative analysis.} Central to our method is the use of BlenderRAG for retrieving relevant up-to-date Blender documentation for writing the code for generating the shape. To demonstrate the influence of this design choice quantitatively, we measure the code complexity of the generated shapes and the code execution error rate. Specifically, we examine two metrics: the number of unique complex Blender operations and the code execution error rate. 

To measure code complexity, we manually categorize Blender operations into two classes: simple and complex. Simple operations include commonly used API calls, while complex ones require advanced multi-step logic. The operation classification allows us to quantitatively measure the sophistication of the code and validates BlenderRAG’s role in enabling code expressivity without increasing error rates. More details and examples of the operation categorization can be found in \cref{sec:additional_quant}. The total error rate measures the cumulative number of errors encountered throughout the entire code generation process. We consider generated code that does not execute successfully as one error. We note that ultimately, all the execution errors are corrected by our system.

\begin{figure}[!t]
    \centering
    \includegraphics[width=\columnwidth]{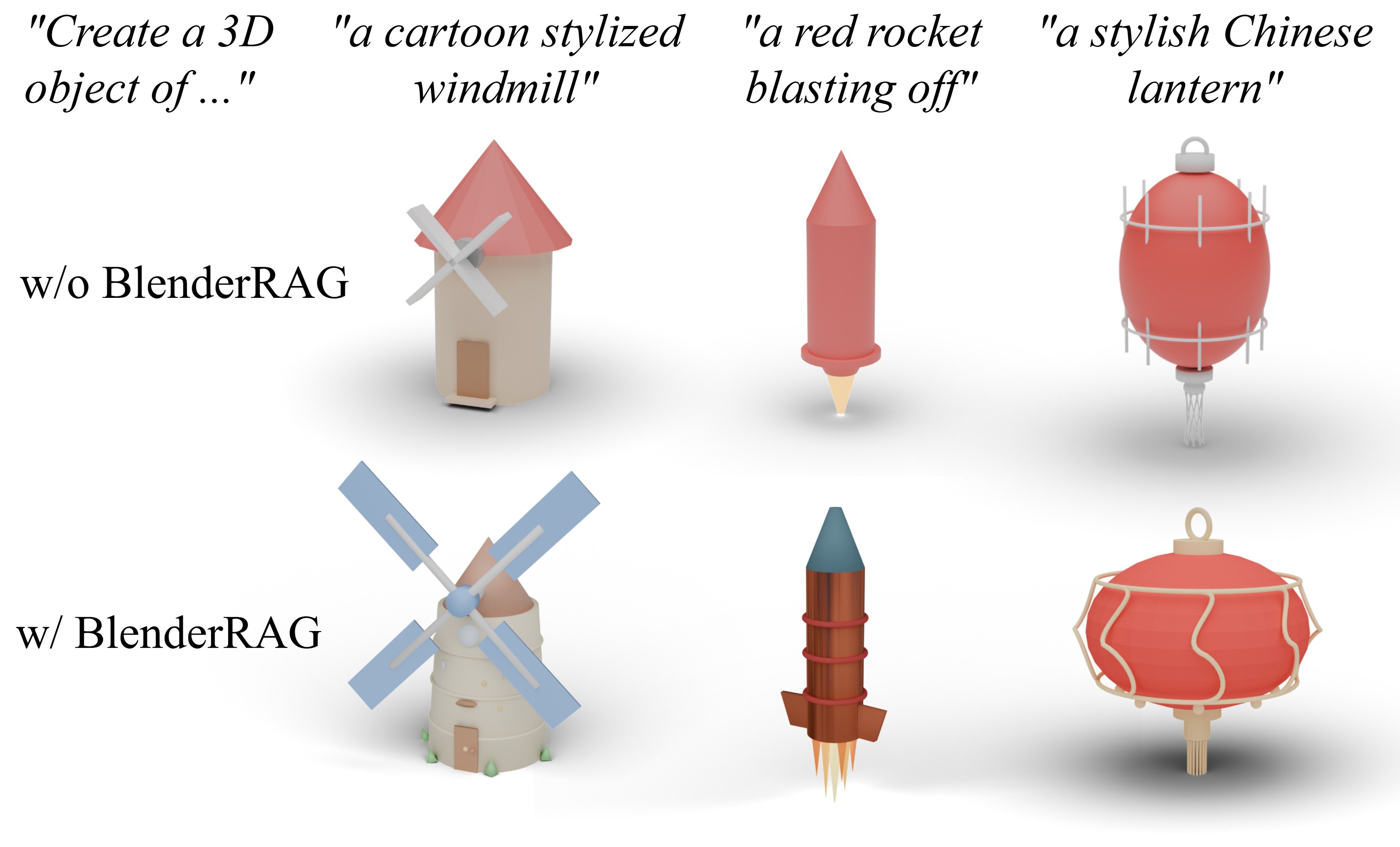}
    \caption{\textbf{The importance of BlenderRAG.} By incorporating a database of Blender API documentation, we enable the coding agent to use more complex functions that improve mesh quality (right) when compared against the lack of such a database (left). For example, the rocket generated with BlenderRAG uses shader nodes to create the reflective material.} 
    \label{fig:worag}
\end{figure}

For the analysis, we curated a diverse set of 17 target objects (\eg hats, animals, lamps). For each object, we generated the mesh with and without the retrieval agent. All scripts were then submitted to an LLM to assess the type and frequency of the operations performed, along with the error rates. A pre-defined prompt that outlines the definitions and examples of simple and complex operations was passed along to the LLM. The scripts were evaluated to determine error rates and assess the type and frequency of operations performed. The results are summarized in \cref{tab:ragflow-error-rates}. A more comprehensive table of the data can be found in \cref{sec:additional_quant}. 

From the table, we observe that BlenderRAG provides substantial benefit in enabling the system to perform complex Blender operations. Without BlenderRAG, the coding agent often defaults to using primitive shapes and simpler constructs, which compromises on mesh quality by skipping more complex, yet effective, Blender operations. With BlenderRAG, there is an observed $5 \times$ increase in the number of complex operations performed, resulting in meshes of higher quality.

\cref{fig:worag} demonstrates the importance of BlenderRAG utilization. Without it, the coding agent resorts to simple solutions and composes the shape out of a few straightforward primitives. However, with BlenderRAG, the generated asset is more expressive. For example, the windmill contains more details, like rods and fins for the turbine, the Chinese lantern contains geometric elements of higher complexity, such as the curved threads, and the rocket has intricate properties, like the glossy material for the body.  

\begin{table}[!t]
\centering
\small
\begin{tabular}{lcccc}
\toprule
\textbf{} & \makecell{\textbf{Complex} \\ \textbf{w/ RAG $\uparrow$}} & 
            \makecell{\textbf{Complex} \\ \textbf{w/o RAG $\uparrow$}} & 
            \makecell{\textbf{Errors} \\ \textbf{w/ RAG $\downarrow$}} & 
            \makecell{\textbf{Errors} \\ \textbf{w/o RAG $\downarrow$}} \\
\midrule
Avg & 5.86 & 1.20 & 2.43 & 3.29 \\
\bottomrule
\end{tabular}
\caption{\textbf{Quantitative results.} We measure code sophistication by counting unique complex function calls with BlenderRAG (Complex w/ RAG) and without (Complex w/o RAG). Using BlenderRAG increases the variety of complex functions used throughout the code without increasing the number of execution errors, leading to higher-quality 3D assets.}
\label{tab:ragflow-error-rates}
\end{table}

Increasing the usage of complex operations has the potential to increase the chance of errors in the code. But surprisingly utilizing BlenderRAG does the opposite, it \textit{reduces} the total error rate across the mesh generation process by $26\%$. This improvement likely stems from the ability of BlenderRAG to provide more accurate and version-specific documentation over multiple refinement iterations. This analysis shows the importance of BlenderRAG in improving shape complexity and reducing error rates, leading to higher quality assets.

\smallskip
\noindent \textbf{Time analysis.} We analyze the processing time of our system by measuring the runtime for each of the three phases, averaged over 33 generations of various objects presented in the paper, including trees, guitar, piano, lamp, and more. The average runtime for the initial creation is about 4 minutes, and for the auto-refinement phases is about 6 minutes. This processing time is required only once at the beginning of the 3D asset creation. Then, further user edits are performed much faster, with an average time of about 38 seconds per edit instruction, enabling a manageable successive manipulation of the asset. %

An interesting behavior we have discovered is that some edits are easier to achieve than others. For example, changing the material (\cref{fig:same-in-diff-mat}) or adding elements to the asset (\cref{fig:iterative}) requires a single user refinement prompt. However, instructions relating to \textit{spatial} location, like adding an ice cream to the character's hand (\cref{fig:humanoid}), require several user follow-up prompts to correct visual errors that the VLM of the critic and verification agent missed. See additional prompt details in the supplementary material (\cref{sec:user_guidance}).

Quantitatively, about 59\% of the user edit examples presented in the paper were obtained with a single edit instruction. In the other cases, 3 to 4 follow-up prompts are needed. Fortunately, since our system \textit{maintains} the generated code during the creation process, the user's edits are implemented as localized code modifications rather than writing the code from scratch, saving substantial processing time. Concretely, instead of regenerating an asset from scratch (running phase one and two) with a new prompt ($\approx$ 10 minutes), we can run iterative refinement (phase 3) on the current generated asset using follow-up prompts that each take about 38 seconds.

\smallskip
\noindent \textbf{Agent ablation.} In \cref{fig:ablation-agent}, we present an ablation study that highlights the importance of each of the agents in creating a high-quality 3D asset. Each agent plays a meaningful role in the system, and mesh quality improves progressively as each is incorporated. As seen in the figure, while the coding agent makes a good first attempt, the planner agent's breakdown of the task allows more parts to be added: a sphere for the hinge of the scissors, more parts at the brush tip, and a piece of fabric for the stocking's hanging loop. The retrieval agent provides Blender code guidance that leads to much more geometrically rich realizations of these added parts: proportional scissor blades, hairs for the brush, and a loop shape for hanging the stocking. Finally, the visual critic provides holistic visual feedback that refines and brings these enhancements together to make for a quality 3D asset.

\smallskip \noindent \textbf{Context sharing ablation.} In our system, the agents have a shared context throughout the entire generation process. Specifically, in the auto-refinement phase, the coding agent has access to the code from the initial creation phase. The shared context is highly beneficial to the 3D generation process, as the coding agent can build on the previous result rather than writing the code from scratch.

\begin{figure}
    \centering
    \includegraphics[width=\columnwidth]{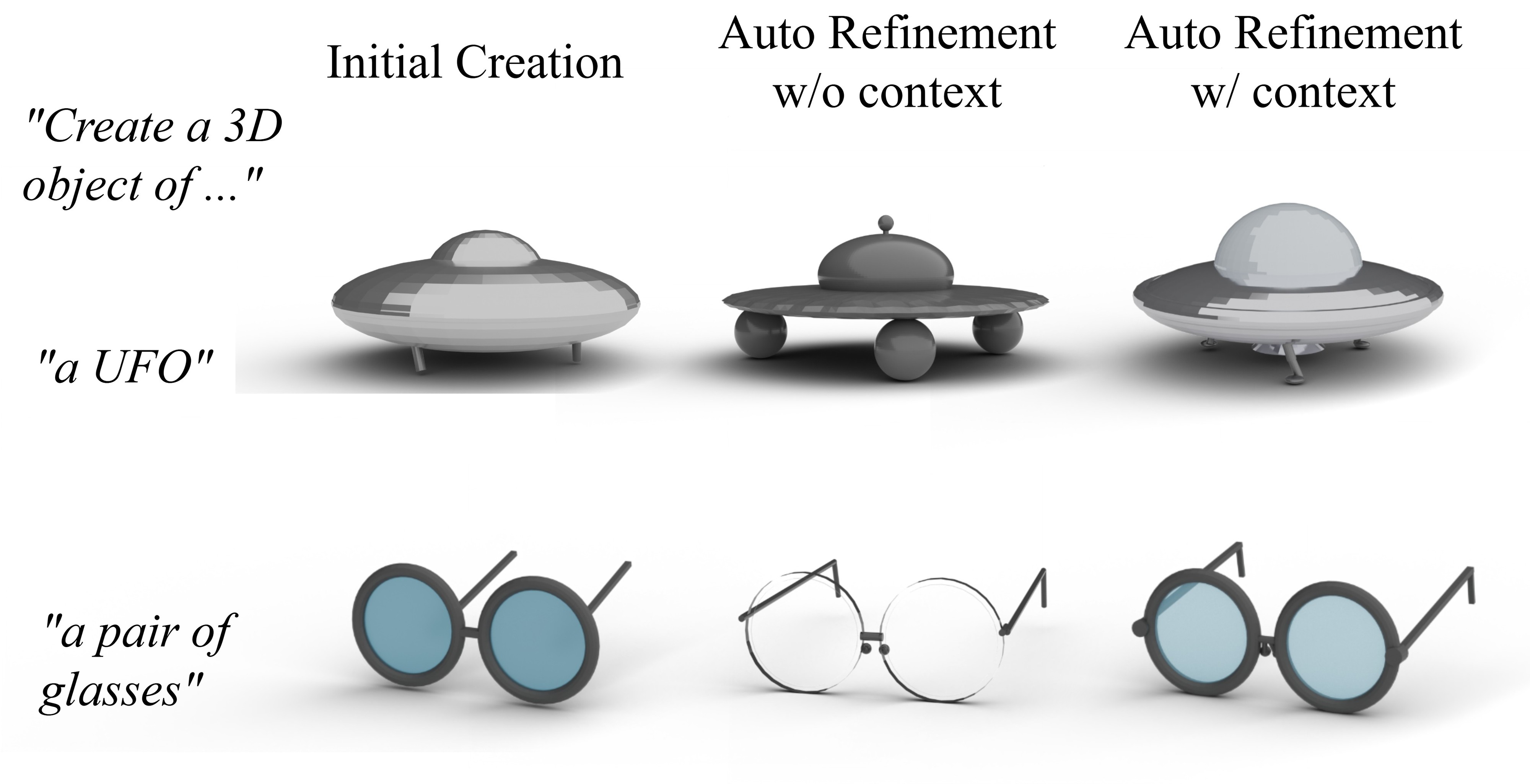}
    \caption{\textbf{Context sharing.} 
    Sharing code context across agents ensures consistent editing and maintenance of the 3D asset throughout the pipeline. In the initial creation phase, an initial mesh is generated (left). If, in the auto-refinement phase, the code generating the initial mesh is not passed to the coding agent, the agent defaults to regenerating the code from scratch, resulting in a different mesh (middle). However, when the previous version of the code is included in the auto-refinement phase, the asset is more faithfully preserved and improved upon (right).}
    \label{fig:context}
\end{figure}

We demonstrate the importance of the shared context in \cref{fig:context}. Without the shared context, the coding agent is still able to create the required asset, such as the UFO and the glasses. While the asset is of high quality, it is substantially different than the initial one. In contrast, the shared context enables the coding agent to refine the asset 
while building on the original attempt, like keeping the initial structure of UFO's landing struts and the frame of the glasses.

\begin{figure}[!t]
    \centering
    \includegraphics[width=\columnwidth]{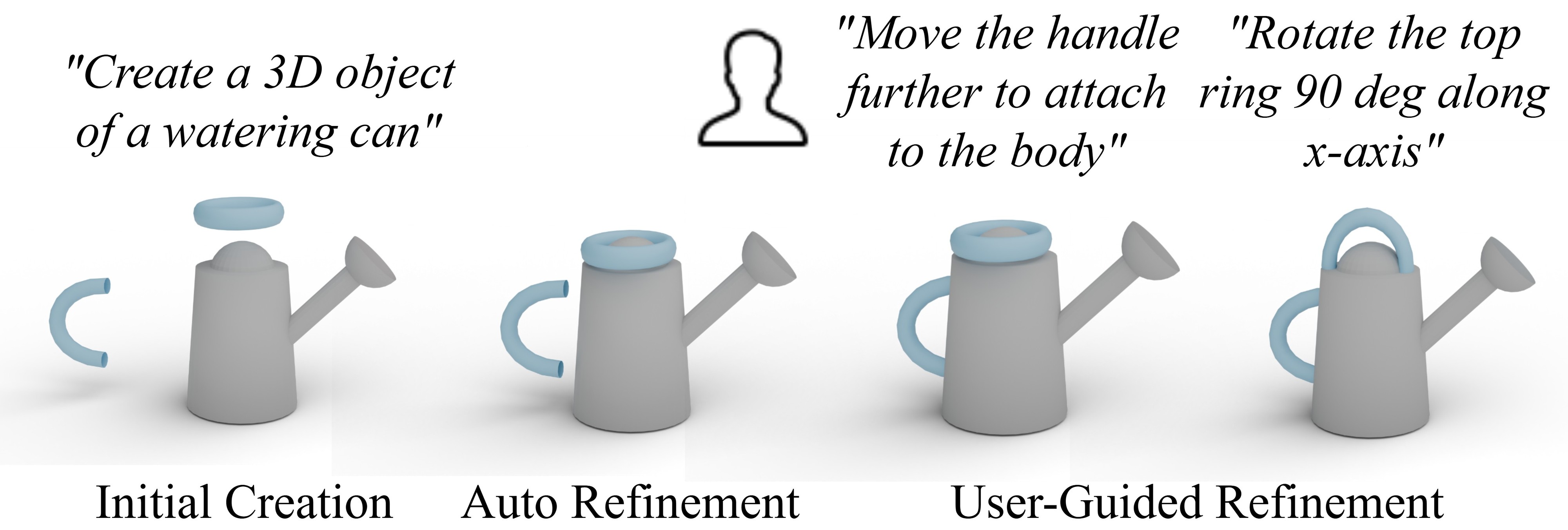}
    \caption{\textbf{Limitations.} \ourmethod{} relies on the accuracy of VLMs to provide automatic visual feedback on the generated shape. However, VLMs may still struggle to accurately identify spatial artifacts, leading to imperfect results after the auto-refinement phase (second column). Our system is able to overcome these limitations by enabling users to supply corrections as additional feedback to the system.}
    \label{fig:limitation}
\end{figure}

\smallskip
\noindent \textbf{Limitations.} Although we incorporate an auto-refinement phase in our method, it may not correct all of the flaws in the initially created shape. For example, the top and side handles of the watering can in \cref{fig:limitation} are disconnected after the initial creation. After the auto-refinement phase, the side handle is closer to the body, but still disconnected, and the top handle has an incorrect horizontal pose. Nonetheless, these issues can be easily remedied via further simple user instructions, showcasing the beneficial editing capability of our method.

\section{Conclusion}
\label{sec:conclusions}
We introduced \ourmethod{}, a multi-agent framework for writing interpretable, modular Python code in Blender that creates 3D assets. Our results showcase the ability of LLMs to perform code-writing tasks that involve accurately describing spatial relationships as seen in \cref{fig:gallery}, such as designing the keys on the piano keyboard, arranging the wheels on a skateboard, or organizing the blades on a windmill. Our system is not constrained by object category or representation type. It flexibly handles a wide range of shape classes and visual styles, generating not only geometry but also textures, colors, and material properties.

A notable feature of our system is the ability to enable users to engage in iterative, dialogue-driven co-creation of 3D models with the system. Beyond one-shot generation, \ourmethod{} supports sequential, high-level user directives that iteratively guide the design process. Users can correct errors, introduce new components, or adjust visual attributes, all through natural language interactions. This interactive loop allows for continuous refinement without regenerating the asset from scratch, preserving coherence and user intent throughout the modeling process.

A particularly compelling aspect of our system is the nature of the generated code: it is interpretable, well-structured, and human-readable, featuring modular variable names and informative comments. Even non-experts can examine the scripts to understand or modify the underlying logic of the generated assets. This promotes transparency, reusability, and creative control, bridging the gap between automation and user authorship.

\section{Acknowledgments}
\label{sec:acknowledgments}

This research was supported by grant \#2022363 from the United States - Israel Binational Science Foundation (BSF), grant \#2304481 from the National Science Foundation (NSF), and gifts from Adobe, Snap, and Google. We thank the members of the 3DL lab for thorough discussions and thoughtful comments. We also extended our gratitude to Sagie Benaim, Rinon Gal, and Yael Vinker for their helpful feedback.

{
    \small
    \bibliographystyle{ieeenat_fullname}
    \bibliography{main}
}
\appendix
\clearpage
\setcounter{page}{1}
\maketitlesupplementary

The following sections provide more information on our multi-agent Blender code-writing 3D modeling system. Section \cref{sec:agent_implementation_details} lists implementation details for the agents, including agent order, instructions, and tools. Section \cref{sec:future_blender_ver} offers an explanation for Blender version adaptability. Section \cref{sec:additional_quant} gives clarifications for the quantitative results from the paper while Section \cref{sec:further_blenderrag} discusses BlenderRAG further in detail. Lastly, Section \cref{sec:user_guidance} provides general guidance on additional user inputs for mesh editing. 

\section{Agent Implementation Details} 
\label{sec:agent_implementation_details}

\subsection{Multi-Agent Coordination} \label{subsec:multi_agent_coordination}
The agents are organized into three teams, one for each phase. Within each team, agent communication is managed by an external orchestrator. It is a central controller that oversees the workflow by managing turn-taking, facilitating agent communication, and handling session termination. Specifically, after each agent's turn, the output is passed to the orchestrator to broadcast the message into the shared context space. The orchestrator then uses a selector function that we defined to determine the agent order. 

\begin{figure*}[h]
    \centering
    \includegraphics[width=\textwidth]{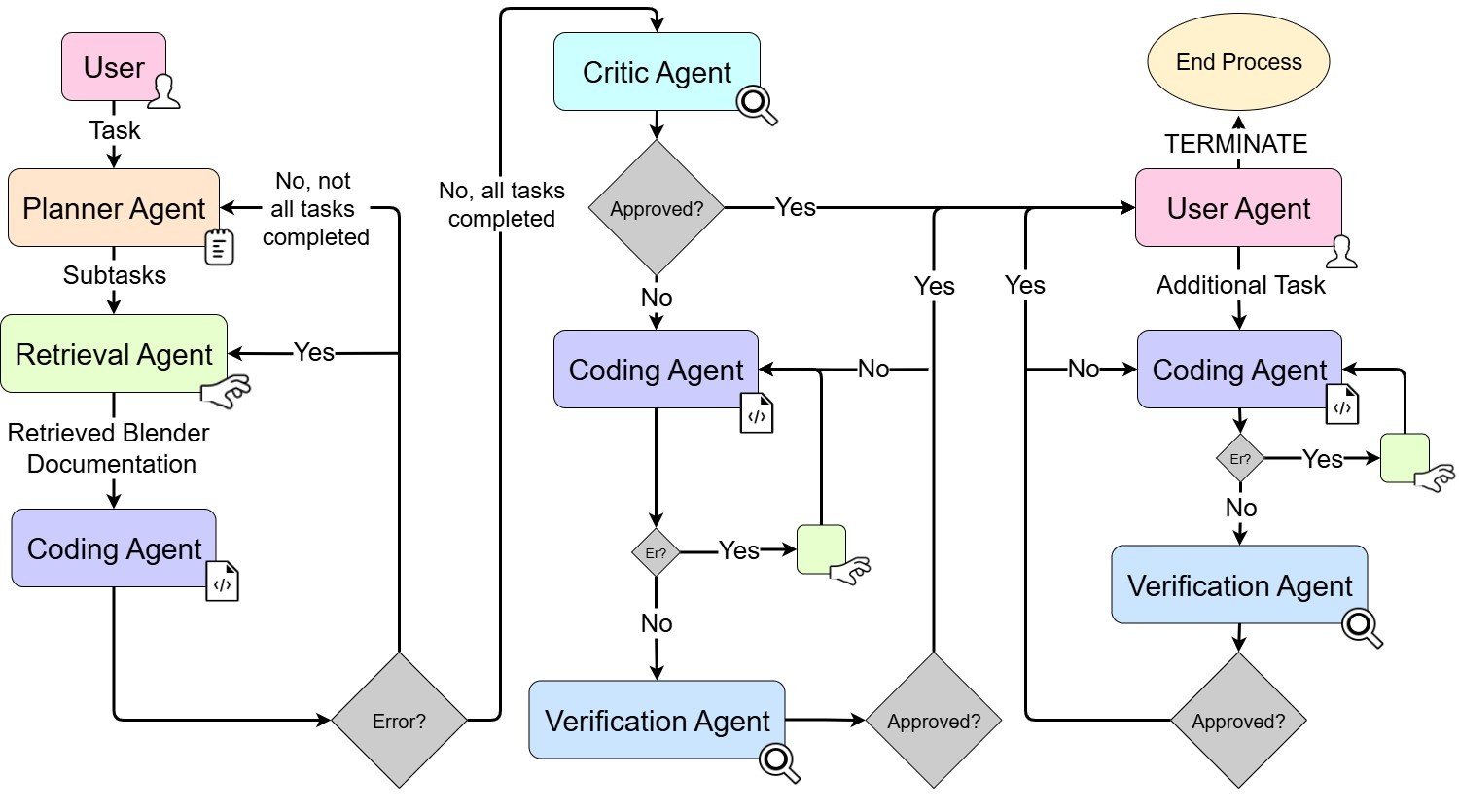}
    \caption{\textbf{Workflow logic.} This diagram is a high-level representation of the agent order logic. The inputs, outputs, and execution conditions are labeled as described in \cref{sec:method}. The white diamond and green box below the coding agent of the second and third columns represent the loop between the retrieval agent and coding agent if an error occurs during code execution.}
    \label{fig:logic}
\end{figure*}

Crucially, the orchestrator also allows all agents to share the same context space, which enables coherent coordination across different agents and the preservation of global contexts across all given tasks. To direct agent ordering of each team, we initialize the orchestrator with a list of agents in that team and explicit conditions that determine when each agent should be invoked. For example, the retrieval agent is invoked when the code agent runs into a code execution error. The full workflow logic can be seen in \cref{fig:logic} The presence of the orchestrator thus ensures that all agents receive the full conversation history and the conversation proceeds in the correct agent order as intended. 

\subsection{Agent Definitions}

\subsubsection{Planner Agent}

The \textit{planner\_agent} uses the \texttt{gpt-4o} model provided by OpenAI. It serves as the central coordinator for the refinement workflow. Its primary responsibility is to break down the high-level user instruction into smaller, manageable subtasks and delegate them to the appropriate agents. Its system prompt is defined as:

\begin{quote}
"You are a planning agent. Your job is to break down complex tasks into smaller, manageable subtasks. Your team members are: \textit{retrieval\_agent}: Retrieves information from the knowledge base. \textit{coding\_agent}: Writes Blender bpy code, implements each subtask one at a time with the help of the additional knowledge from the \textit{retrieval\_agent}. You only plan and delegate tasks --- you do not execute them yourself.

The \textit{retrieval\_agent} will provide you with additional info about the official bpy documentation implementation. The \textit{coding\_agent} should refer to the official bpy documentation implementation when writing the code to avoid errors. When the code execution fails, the \textit{coding\_agent} should use the \textit{retrieval\_agent} to find solutions to the errors.  

\textit{Coding\_agent} must provide code as an argument to use the \texttt{execute\_code\_tool}.

When \textit{coding\_agent} encounters an error, it should use the \textit{retrieval\_agent} by calling the \texttt{retrieve\_information\_tool} to find solutions to the errors.

When assigning tasks, use this format: \texttt{<agent>:<task>}. After all tasks are complete, summarize the findings and end with \texttt{"COMPLETE"}."
\end{quote}

\subsubsection{Retrieval Agent}

The \textit{retrieval\_agent} uses the \texttt{gpt-4o} model from OpenAI. It retrieves relevant information from a BlenderRAG knowledge base using the \texttt{retrieve\_information\_tool}. This agent is typically invoked when the \textit{code\_agent\_refine} encounters an error or needs clarification about a particular Blender API usage. Its system prompt is defined as:

\begin{quote}
"You are a helpful assistant who can retrieve information from the knowledge base. If the \textit{code\_agent} encounters an error, you must use the \texttt{retrieve\_information\_tool} to retrieve the information from the knowledge base. Execute the \texttt{retrieve\_information\_tool} with the error message as the argument to retrieve the information from the knowledge base."
\end{quote}

\subsubsection{Coding Agent}

The \textit{code\_agent\_refine} uses the 
\texttt{claude-3-7-sonnet} model from Anthropic. It is responsible for editing and executing Blender \texttt{bpy} scripts. Its system prompt is defined as:

\begin{quote}
"You are a helpful assistant who can write Blender bpy code. You will be given a task, and you will need to write the code to complete the task. You must use the \texttt{execute\_code\_tool} to execute the code."
\end{quote}

\subsubsection{Critic Agent}

\begin{figure*}[h]
    \centering
    \includegraphics[width=\textwidth]{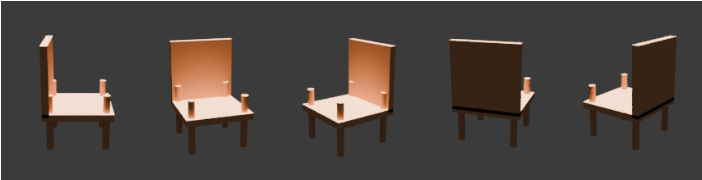}
    \caption{\textbf{Render function example.} A chair from the initial creation phase is rendered from five angles that together capture a holistic overview of the object. These renders are then passed into an external VLM by the critic agent for evaluation. We use the eevee rendering mode in Blender to reduce the mesh generation time.}
    \label{fig:renders}
\end{figure*}

The \textit{critic\_agent} uses the \texttt{gpt-4o} model and visually critiques the Blender scene based on the user’s original prompt. It utilizes the \texttt{critique\_scene\_tool}, which renders the scene from multiple angles and submits it to a vision-language model (Gemini) to identify mismatches or disconnected components. An example of the five renders is shown in \cref{fig:renders}. Its system prompt is defined as:

\begin{quote}
"You are a helpful assistant who can critique the scene. You will use the \texttt{critique\_scene\_tool} to critique the scene. You will be given the target prompt, you need to pass the target prompt as an argument to the \texttt{critique\_scene\_tool}."
\end{quote}

\subsubsection{Verification Agent}

\begin{figure}[h]
    \centering
    \includegraphics[width=\columnwidth]{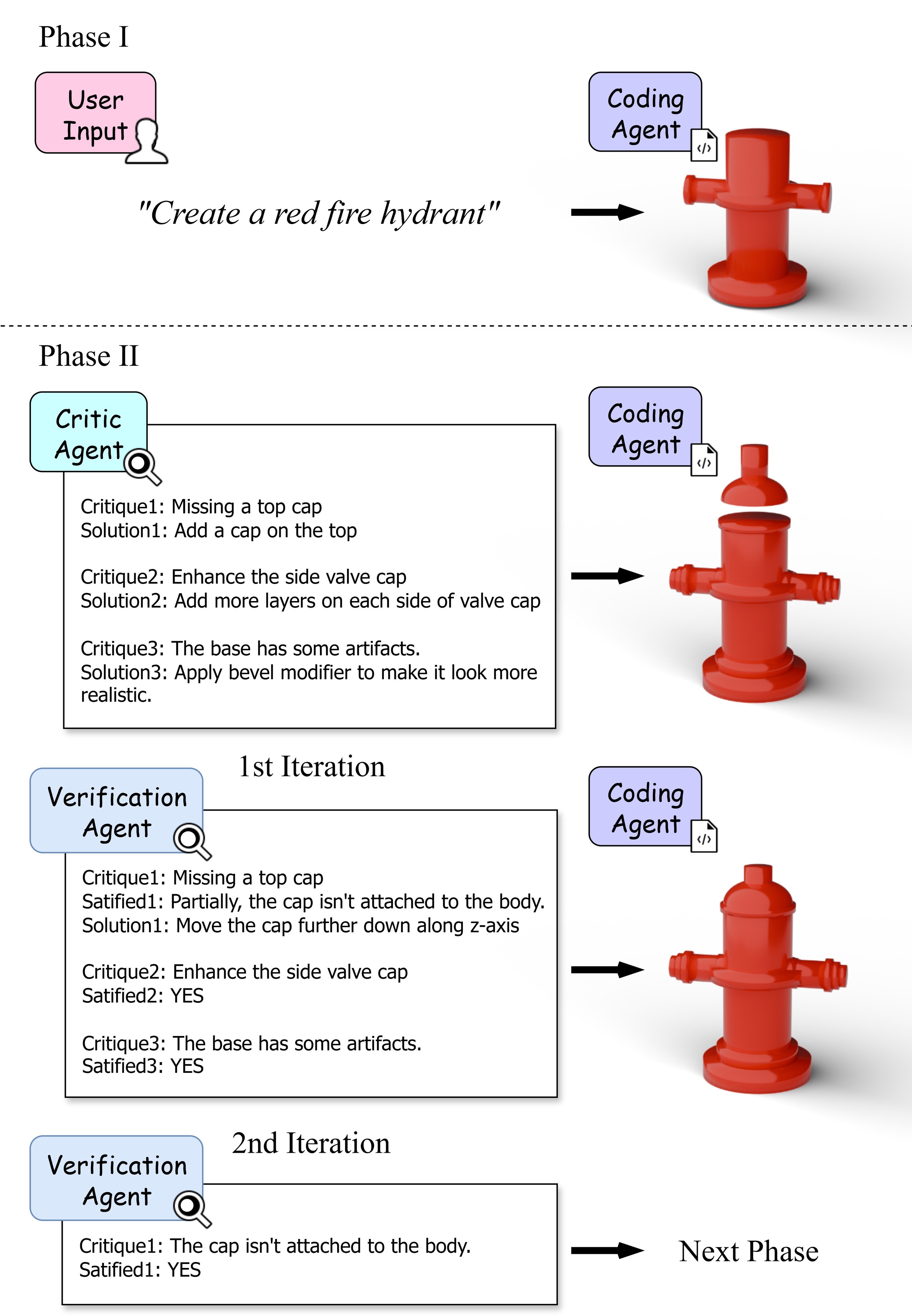}
    \caption{\textbf{Verification agent.} The coding agent in the auto-refinement phase may not always address all the issues on its first attempt, hence requiring the presence of a verification agent to check the implementation. In the figure, only the second and third critiques were addressed by the coding agent in the first iteration, thus requiring a second iteration by the coding agent to address the first critique.}
    \label{fig:verification}
\end{figure}

The \textit{verification\_agent} also uses the \texttt{gpt-4o} model and determines whether the updated Blender scene satisfies all critiques provided by the \textit{critic\_agent}. It uses the \texttt{verify\_scene\_tool}, which re-renders the scene and evaluates it against the saved critique list via the Gemini vision-language model. An example of how the verification agent works can be seen in \cref{fig:verification}. Its system prompt is defined as:

\begin{quote}
"You are a helpful assistant who can verify the scene. You must use the \texttt{verify\_scene\_tool} to verify the scene."
\end{quote}

\subsubsection{User Proxy Agent}

\begin{figure*}[h]
    \centering
    \includegraphics[width=\textwidth]{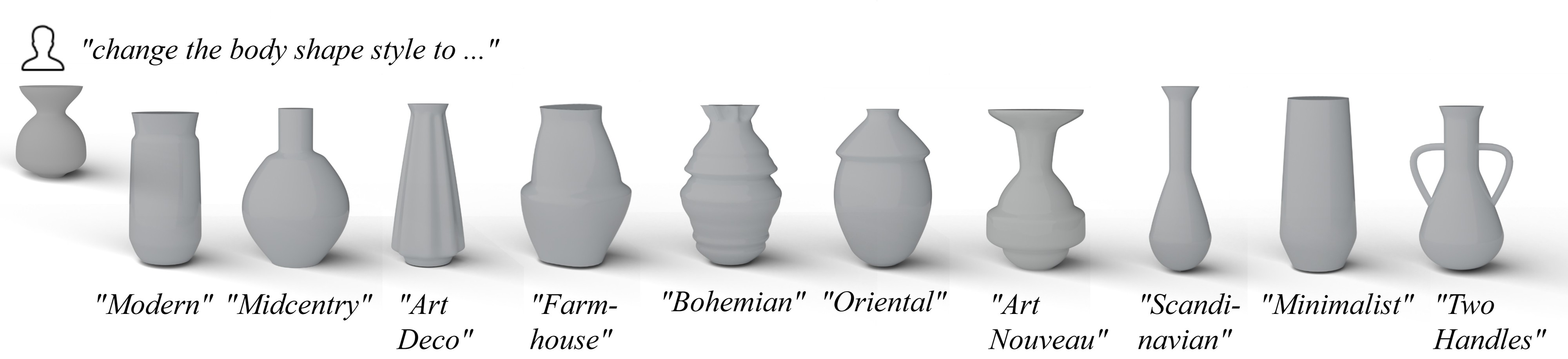}
    \caption{\textbf{Geometry Refinement.} Given a an initial mesh produced by our system, LL3M may additionally modify the geometry based on user instructions in the form of text prompts. Our method applies flexible geometric edits, like changing the shape’s outline (“Midcentury”, “Scandinavian”), creating surface details (“Art Deco”, “Bohemian”), and even adding new parts (“Two Handles”).}
    \label{fig:same-in-diff-out}
\end{figure*}

The \textit{user\_proxy\_agent} is a special interface agent that does not use any language model. It acts as a bridge between the system and a human operator, allowing interactive feedback during the refinement process. It accepts text directly from the user. This instruction is passed into the shared context to be further processed by other agents to modify the 3D asset. The updated code is then executed in Blender. The user may then continue with additional instructions given the visual updates in Blender from the previous iteration. This loop ensures that the final output aligns more closely with user intent but also allows for creative geometric edits like \cref{fig:same-in-diff-out}.

\section{Future Blender Versions}
\label{sec:future_blender_ver}
In our work, we use BlenderRAG for a specific Blender version. However, BlenderRAG can be further customized by adding future Blender versions' documentation (or even custom documentation knowledge bases) into the RAG database, letting users work with the most up-to-date Blender API independent of the code agent's LLM and its knowledge of Blender up to the training cutoff.

\section{Additional Quantitative Analysis Details}
\label{sec:additional_quant}

To evaluate the effectiveness of BlenderRAG, We looked at 34 scripts of \emph{bpy} code across 17 diverse objects such as Chinese Lantern, Crocodile, Windmill, and more. The scripts were collected and passed into \texttt{gpt-4o} along with a prompt defining simple/ complex Blender operations for counting. Simple operations include one-line or commonly used API calls that do not require deeper manipulation of Blender's data structures. Examples include setting the active object into a different mode (\eg \texttt{bpy.ops.object.mode\_set}) or adding a primitive (\eg \texttt{bpy.ops.mesh.primitive\_uv\_sphere\_add}). We also consider operations such as appending materials to objects, setting vertex groups, or basic scaling transformations as simple.

\begin{table}[h!]
\centering
\small
\begin{tabular}{lcccc}
\toprule
\textbf{} & \makecell{\textbf{Complex} \\ \textbf{w/ RAG}} & 
                   \makecell{\textbf{Complex} \\ \textbf{w/o RAG}} & 
                   \makecell{\textbf{Errors} \\ \textbf{w/ RAG}} & 
                   \makecell{\textbf{Errors} \\ \textbf{w/o RAG}} \\
\midrule
Bottle           & 4  & 0  & 3 & 3 \\
Bowtie           & 3  & 1  & 4 & 3 \\
Brush            & 8  & 0  & 2 & 3 \\
Bunny            & 6  & 1  & 3 & 4 \\
Castle           & 4  & 2  & 4 & 3 \\
Cat              & 6  & 2  & 3 & 4 \\
Chair            & 5  & 0  & 2 & 5 \\
Lantern          & 5  & 0  & 3 & 5 \\
Crocodile        & 6  & 2  & 3 & 3 \\
Dragon           & 9  & 2  & 2 & 3 \\
Fish             & 7  & 3  & 3 & 2 \\
Flag             & 4  & 1  & 2 & 3 \\
Hat              & 8  & 1  & 3 & 4 \\
Lamp             & 5  & 2  & 0 & 5 \\
Rocket           & 2  & 0  & 0 & 0 \\
Windmill         & 5  & 2  & 2 & 2 \\
\midrule
\textbf{Avg}     & 5.86 & 1.21 & 2.43 & 3.29 \\
\bottomrule
\end{tabular}
\caption{\textbf{Quantitative results.} We looked at 30 scripts in total (with and without RAG) and counted the number of simple/ complex Blender operations and errors. These scripts, along with a prompt defining what simple/ complex operations are and with examples, are fed into GPT-4o for evaluation.}
\label{tab:supp-error}
\end{table}

In contrast, complex operations involve advanced functionality that requires multi-step logic, such as manual mesh editing via \texttt{bmesh}, shader node construction, geometry node setups, and procedural geometry composition or manipulation of object data. Examples include operations that create vertices and faces (\eg \texttt{bmesh.new()}, modify shaders like (\eg \texttt{bsdf.inputs['Base Color'].default\_value}), or directly adjust vertex coordinates. These complex operations allow for meshes of higher quality and detail. The full data table can be seen in \cref{tab:supp-error}.

\section{Further discussion on BlenderRAG}
\label{sec:further_blenderrag}

In \cref{tab:ragflow-error-rates}, we observed that BlenderRAG reduced the execution error rate, but the reduction is relatively mild. This less-than-expected improvement can be due to a combination of factors: BlenderRAG may still miss some critical details in the first attempt, and without BlenderRAG, the system often simplifies its solutions to avoid encountering errors, thus reducing the number of failed attempts at the cost of output quality.

For some cases without BlenderRAG, the coding agent often "solves" the problem by deleting the lines that caused the error. This observation supports how BlenderRAG enables more sophisticated code without increasing error rates, since it retrieves examples to fix issues instead of deleting buggy code lines. 

\begin{figure}[h]
    \centering
    \includegraphics[width=\columnwidth]{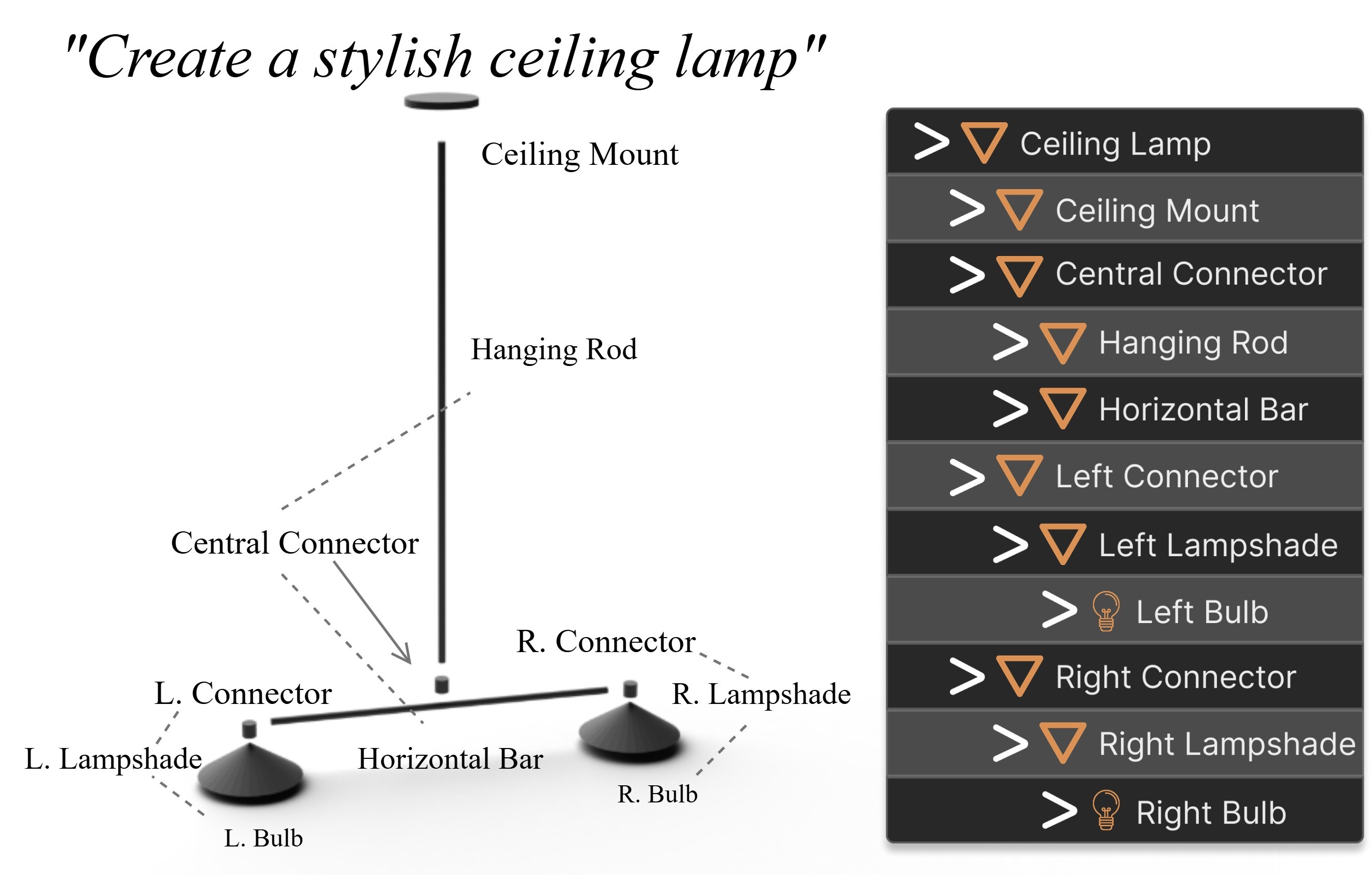}
    \caption{\textbf{Hierarchical scene graph.} The coding agent establishes logical parenting relationship when the input prompt explicitly asks for such a relationship. Doing so generates shapes with a human-readable hierarchical structure by creating parent-child relationships between parts within the scene. This enables scene graph behavior in Blender, where transformations applied to a parent propagate to its children. Each part in the graph is also assigned a meaningful semantic name.}
    \label{fig:parenting}
\end{figure}

We also observed the ability of the coding agent, with the aid of BlenderRAG, to construct parenting relationships. While the scripts that were generated using RAGFlow have a much higher frequency of using complex operations, this is not limited to code. When explicitly defined in the user input prompt, we noticed that the coding agent is able to establish logical parenting relationships between 3D assets in Blender (\eg in \cref{fig:parenting}, the \textit{left connector} is the parent of \textit{left landshape and bulb}), allowing intuitive control and structural coherence. This also grants greater modularity for the user to understand and modify the created mesh as they wish.

\section{Additional Details for User Guided Refinement}
\label{sec:user_guidance}

There are cases where a complex mesh requires multiple user prompts in the user guided refinement phase in order to achieve high-quality results. This is often due to a combination of LLM hallucination, imprecise prompting, and imperfect spatial awareness of the VLM. For example, in the case of \cref{fig:humanoid}, the creation of the humanoid figure was guided by the following series of prompts, in sequence:

\begin{enumerate}
    \renewcommand{\labelenumii}{\theenumi.\arabic{enumii}}
    
    \item \textbf{Create a mini cartoon character}
    
    \begin{enumerate}
        \item Move the eye pupils outward along the $x$-axis.
        \item Move the hands slightly backward along the negative $x$-axis to attach them to the arms.
    \end{enumerate}
    
    \item \textbf{Add a blonde wig on the head}
    
    \begin{enumerate}
        \item Scale the wig smaller.
        \item Move the wig upward along the $z$-axis to place it on top of the head.
    \end{enumerate}
    
    \item \textbf{Add a pair of glasses to the face}
    
    \begin{enumerate}
        \item Rotate the glasses $90^\circ$ around the $y$-axis so they attach to the legs.
        \item Move the glasses along the negative $x$-axis to position them closer to the face.
        \item Move the glasses upward along the $z$-axis to align with the eyes.
    \end{enumerate}
    
    \item \textbf{Add a sprinkled ice cream to the left hand}
    
    \begin{enumerate}
        \item If the ice cream is not attached to the cone, move it slightly downward along the $z$-axis.
        \item Add more sprinkles and make them larger and denser around the ice cream.
        \item If the ice cream is on the character’s arm, move it slightly along the $y$-axis to place it in the hand.
    \end{enumerate}
    
    \item \textbf{Make the character sit down and eat the ice cream with both hands}
    
    \begin{enumerate}
        \item If the legs are overlapped, move them outward slightly to separate them.
        \item Rotate only the arms so both hands are positioned to hold the ice cream.
        \item Rotate the ice cream along the $z$-axis so the camera can see the sprinkles on top.
    \end{enumerate}
\end{enumerate}

\end{document}